# Adaptive Streaming in Interactive Multiview Video Systems

Xue Zhang, Laura Toni, *Member, IEEE,* Pascal Frossard, *Fellow, IEEE,*
Yao Zhao, *Senior Member, IEEE,* and Chunyu Lin

***Abstract*—Multiview applications endow final users with the possibility to freely navigate within 3D scenes with minimum-delay. A real feeling of scene navigation is enabled by transmitting multiple high-quality camera views, which can be used to synthesize additional virtual views to offer a smooth navigation. However, when network resources are limited, not all camera views can be sent at high quality. It is therefore important, yet challenging, to find the right tradeoff between coding artifacts (reducing the quality of camera views) and virtual synthesis artifacts (reducing the number of camera views sent to users). To this aim, we propose an optimal transmission strategy for interactive multiview HTTP adaptive streaming (HAS). We propose a problem formulation to select the optimal set of camera views that the client requests for downloading, such that the navigation quality experienced by the user is optimized while the bandwidth constraints are satisfied. We show that our optimization problem is NP-hard, and we therefore develop an optimal solution based on the dynamic programming algorithm with polynomial time complexity. To further simplify the deployment, we present a sub-optimal greedy algorithm with effective performance and lower complexity. The proposed controller is evaluated in theoretical and realistic settings characterized by realistic network statistics estimation, buffer management and server-side representation optimization. Simulation results show significant improvement in terms of navigation quality compared with alternative baseline multiview adaptation logic solutions.**

***Index Terms*—HAS, adaptation logic, multiview navigation, content characteristics.**

## I. Introduction

WITH recent advances in interactive and immersive video technologies, multiview (MV) and 360-degree videos [1] applications are gaining ever-increasing popularity, e.g., virtual reality (Microsoft HoloLens [2] and Oculus Rift [3]). Their advent has enabled users with the ability to freely navigate within a 3D scene via images captured from multiple cameras. In the case of MV, this is possible due to the free-viewpoint technology, where a virtual viewpoint can be synthesized at the decoder via depth-image-based rendering (DIBR) [4] using both the texture and the depth maps of camera views, namely anchor views. The quality of the synthesized viewpoints generally increases with both the quality of the anchor views and the similarity or proximity between the anchor views and the synthesized views. High-quality rendering of the scene is thus expected to require multiple high-quality camera views, which is however not always feasible with limited network resources. It therefore becomes essential to properly select the camera views to transmit to users.

HTTP adaptive streaming (HAS), the defacto technology for video streaming over the Internet, provides an ideal framework to address this challenge. The key concept underpinning HAS is to pre-encode each video at different encoding bitrates and/or resolutions (i.e., representations) and store them at the main server. Each client then selects the best representation to download over time based on his requirements, network bandwidth, playback buffer level, etc [5]. This downloading optimization is called adaptation logic. How to properly perform this optimization is a very challenging task and it has been under intense research lately. Most of research efforts to date have however focused on classical video streaming applications, while a few works have studied the adaptation logic for HAS-based MV streaming. The works in [6]–[8], for example, have taken into account the importance of seamless view switching upon request in MV adaptive streaming. In particular, Hamza and Hefeeda [8] have highlighted that each component of anchor views contributes differently to the quality of the synthesized view in their free viewpoint streaming optimization. However, the limitation of these works is that the scene characteristics are neglected, which are definitely critical to determine the best tradeoff between coding and virtual view artifacts.

Therefore, in this paper, we propose a novel optimization of adaptation logic for MV interactive users with proper consideration of video characteristics and user interactivity. In more details, we consider the scenario of MV video sequences stored at the main server of the service provider (e.g., Netflix, YouTube). Each view corresponds to a sequence of texture images and depth maps captured by a given camera. Each view is pre-encoded into different representations. Each representation is then decomposed into temporal video segments (usually 2s long) and stored at the server. Then the client requests the best set of representations for the current video segment based on both its level of interactivity and the available

This work was supported in part by the National Key R&D Program of China under Grant 2016YFB0800404, the National Natural Science Foundation of China under Grant 61532005, 61332012, 61772066, the China Scholarship Council, and supported in part by the Swiss National Science Foundation under the CHIST-ERA project CONCERT (A Context-Adaptive Content Ecosystem Under Uncertainty, No. FNS 20CH21_151569).

X. Zhang, Y. Zhao, and C. Lin are with the Institute of Information Science, Beijing Jiaotong University, Beijing 100044, China, and also with the Beijing Key Laboratory of Advanced Information Science and Network Technology, Beijing 100044, China (e-mail: xuezhang@bjtu.edu.cn; yzhao@bjtu.edu.cn; cylin@bjtu.edu.cn).

L. Toni is with Department of Electronic & Electrical Engineering, University College London (UCL), London WC1E 7JE, U.K. (e-mail: l.toni@ucl.ac.uk).

P. Frossard is with the Signal Processing Laboratory (LTS4), Ecole Polytechnique Federale de Lausanne (EPFL), Lausanne 1015, Switzerland (e-mail: pascal.frossard@epfl.ch).

bandwidth. Given that the client is interested in a specific navigation window (range of consecutive virtual views that can be displayed by the client), the best set of representations is defined as the one that permits to effectively reconstruct the entire navigation window at the highest quality, subject to the channel constraints. To achieve this goal, we provide a new problem formulation to jointly optimize the subset of camera views and their encoded video bitrates that should be requested by the client, among the ones available at the server. The proposed optimization leads to an optimal solution that takes into account both coding and virtual synthesis artifacts that affect the navigation quality. We show that our optimization problem is NP-hard [9], and we propose an optimal solution based on dynamic programming (DP) algorithm with polynomial time complexity. To further reduce computational complexity, a suboptimal greedy algorithm is also proposed. We compare the adaptation logic strategy with baseline algorithms from the literature. Simulation results show substantial gains in terms of navigation quality under different streaming scenarios. This means that the proposed optimization framework is able to find the right combination of representations that exploits at best the available resources for the considered client. This reflects into a better usage of the available network resources and into higher satisfaction of the final users.

In summary, the contributions of this paper are the following:

(i) We formulate a novel navigation-aware representation selection problem to jointly optimize the set of camera views and their encoded video bitrates for MV adaptive streaming, where the view-switching delay is minimized by considering a navigation window for the users.

(ii) We develop an optimal solution based on DP algorithm with polynomial time complexity. We then propose a suboptimal greedy algorithm in order to reduce the computational complexity of the representation selection algorithm in more realistic systems.

(iii) We provide extensive simulations under different streaming scenarios and show the gains (in terms of navigation quality) of the proposed algorithm with respect to baseline adaptation logics. Realistic network throughput estimation, buffer control and server-side representation optimization are incorporated in the realistic settings.

The remainder of this paper is organized as follows. Related works are reviewed in Section II. In Section III, we formulate the MV representation selection optimization problem. Our optimal and approximate solutions are described in Section IV. Adaptive streaming framework and simulation results are provided in Section V and Section VI, respectively. Finally, the concluding remarks are given in Section VII.

## II. Related Works

In the literature, most of the research efforts on HTTP adaptive streaming have focused on monoview videos [10]–[13], while we directly consider the MV/interactive scenarios. The main difference is that the quality experienced by interactive MV users is highly dependent on particular factors like view synthesis artifacts and switching delays. Optimal coding structure [14]–[17] or resource allocation [18]–[20] for MV and 360-degree videos have been proposed to improve interactive media services in the case of resource-constrained networks. However, they are not carried out on HAS systems. HAS-related MV streaming has been studied from a provider perspective aiming at how to efficiently encode and stream the camera views corresponding to the users' viewpoint [6], [21], and how to optimize the representations to store at the server [22], [23]. All these works however are under a simplified assumptions on the client control strategy. Our work is rather complementary to [22] since we focus on optimizing the adaptation logic strategy at the client side. In the following, we only provide the advances that are mostly related to our work and we comment on the recent ones in HAS systems, with a deep focus on MV systems.

Zhang et al. [24] propose a priority-based adaptive scheme for HAS-based MV streaming, in case that video streams of different scenes (views) are simultaneously transmitted over the bandwidth-constrained network by the server. The encoded quality bitrate is scheduled according to the priority of each view that the client requests. Rather than viewing concurrent streams, the problem of switching from stream to stream is addressed in [25]. A prefetching policy is a key component in which cameras and associated bitrates are adapted based on the steam switching probabilities and the current bandwidth constraints. However, neither of them considers to make up the current viewpoint from multiple parallel streams. In [7], the delay between the view-switching request and the target-view rendering is minimized by employing a buffer occupancy control, parallel streaming, and a server push policy. The above works, however, do not consider the depth information to best support view synthesis.

With the idea of transmitting depth information to aid the generation of virtual views, few related works in the literature have focused on HAS systems for MV streaming [8], [26]–[30]. In [26], the authors introduced a DASH-based multi-modal 3D video streaming system referred to as OmniViewer, which allows the client to view the 3D video from arbitrary sides. However, they focus on the system architecture overview and do not propose any optimization for rate adaptation. In [27], a DASH-based system for stereoscopic 3D videos is presented, but mainly focuses on automatic depth adjustments by taking into account both the content and the display type. A cloud-assisted and DASH-based interactive MV system is proposed in [28], where virtual views to be rendered at either the server or the client based on network conditions and the cost of the cloud-based server. However, they ignore the interaction latency, which nevertheless becomes a crucial element in delay sensitive applications. With free viewpoint video streaming, the work in [29] designs a rate adaptation architecture for texture and depth components using the rate-distortion (R-D) based allocation to maximize the quality of rendered virtual views. Similar to their systems, in our work we target on multi-view-plus-depth videos while virtual views are rendered via DIBR. Unlike their systems, however, we consider a navigation window to minimize the view-switching delay, rather than a single viewpoint with adaptive bitrate.

The most closely related to our work are two adaptation logics proposed in [30] and [8]. In [30], the client adjusts the downloading bitrate by varying the number of anchor views but under the constraint of equal encoded video bitrate for all camera views. Equal bitrate across views is a limiting constraint in multiview systems [19], [31]. Moreover, in their adaptive MV streaming system, multiple views are jointly encoded with the 3D video extension of HEVC (3D-HEVC), which does not offer much flexibility in term of both view and rate adaptation. A two-step rate adaptation approach for free-viewpoint video streaming is further proposed in [8], where the reference views are chosen and then the optimal bitrate for each selected view is determined. While this non-joint optimization neglects the scene characteristics, which does not provide the complete and optimal solution for interactive MV streaming. We rather propose to optimize jointly the reference views and the encoding bitrates, which permits to find better tradeoff and to reach significantly higher performance in most settings by carefully considering the video content characteristics, the user behavior and the network availability altogether.

TABLE I
MAIN NOTATIONS

| Symbol | Definition |
|---|---|
| $\mathcal{V}, v \in \mathcal{V}$ | set of camera views and specific camera view $v$, respectively |
| $\mathcal{R}, r \in \mathcal{R}$ | set of encoding bitrates and specific bitrate $r$ that camera views encoded at, respectively |
| $\mathcal{U}, u \in \mathcal{U}$ | set of viewpoints and displayed viewpoint $u$, respectively |
| $\mathcal{N}, n \in \mathcal{N}$ | set of video segments and specific video segment $n$, respectively |
| $w(u) = [U_L, U_R]$ | navigation window centered in $u \in \mathcal{U}$ with lateral viewpoints $U_L$ and $U_R$ |
| $\Delta$ | minimum space between two adjacent viewpoints |
| $\tau$ | video segment duration |
| $\mathcal{L}$ | set of representations stored at the server |
| $\mathcal{L}_d \subseteq \mathcal{L}$ | set of representations at the decoder |
| $D(\mathcal{L}_d, w)$ | distortion experienced by a user navigating in the window $w$ |
| $d_u(v_i, r_i, v_j, r_j)$ | distortion of viewpoint $u$ synthesized from the representations $(v_i, r_i, v_j, r_j)$ |
| $C$ | downloading bandwidth constraint |

A theoretical formulation for the representation selection optimization has been studied in [32]. This paper is a nontrivial extension, with three important differences: 1) Unlike [32] where heterogeneous users are characterized by a theoretical model, we explicitly perform the deployment of the proposed solution in realistic scenarios with realistic network statistics as well as playback buffer integrated. Furthermore, the server-side representation optimization is employed, where the storage cost as well as different types of users in terms of access link capacity and interactivity level are considered. 2) In order to reduce the computational complexity of the representation selection algorithm in realistic settings, we further develop a suboptimal greedy solution with effective performance by fully taking into account video characteristics. 3) We finally show the significant gains with respect to state-of-the-art works in reality, e.g., the activation of joint coding process in [30], rather than a modified version in [32].

## III. Problem Formulation

In this section, we describe first the system model for on-demand MV adaptive streaming. Then, we define the distortion experienced by a client while navigating through the scene. Finally, we formulate the optimization problem for the MV representation selection at the client. Fig. 1 presents an overview of our problem formulation framework, while Table I summarizes the main notations used in the following.

### A. System model

We consider the MV-based adaptive streaming system depicted in Fig. 2. Let $\mathcal{V} = \{1, 2, 3, \cdots, |\mathcal{V}|\}$ be the set of camera views acquiring the scene. Each view acquires both texture images and depth maps. Each texture image is pre-encoded into different *representations*. Without loss of generality, we consider one spatial resolution and multiple encoding bitrates for each multiview video sequence. Let $\mathcal{R}$ be the set of encoding bitrates for each camera view, and

$$\mathcal{L} = \{(v_i, r_i)\}_i, \text{with } v_i \in \mathcal{V}, r_i \in \mathcal{R} \tag{1}$$

be the set of multiview video representations stored at the main server and made available to clients [1]. Each $i$ of $\mathcal{L}$ is a complete representation set for a given camera view $v_i$, see part A in Fig. 1. The pair $(v_i, r_i)$ identifies a representation that the video sequence captured at camera view $v_i$, whose texture map is encoded at bitrate $r_i$. Since accurate depth information has high importance for the consistency in view synthesis yet relatively low encoding bitrate cost, depth maps are encoded once but at high quality. A constant encoding bitrate is therefore added to that of texture images. The resulting representations $\mathcal{L}$ are divided into video segments with equal playback duration $\tau$ (typically 2s long), which are stored at the main server.

At the client side, the set of viewpoints that can be potentially displayed is denoted by $\mathcal{U} = \{1, 1 + \Delta, 1 + 2\Delta, \cdots, 2, \cdots, |\mathcal{V}|\}$, where $\Delta \in [0, 1)$ is the minimum space between two adjacent virtual viewpoints. We consider that $u \in \mathcal{U}$ identifies any either virtual viewpoint or camera view that can be displayed during the navigation. Any virtual viewpoint $u$ can be rendered using a pair of left and right reference view $v_L$ and $v_R$, $v_L < u < v_R$ and $v_L, v_R \in \mathcal{V}$, via a DIBR technique [4].

For each navigation segment, the client sends a downloading request to the server. The index $n \in \mathcal{N}$ represents the video segment displayed by the client, while $n + \ell$ is the video segment to be downloaded. There is therefore a mismatch between the downloading and displaying time, which is equal to $T = \ell\tau$ seconds. Assuming that the last view displayed in $n$ is denoted by $u \in \mathcal{U}$, and that $\rho$ is the maximum speed (in views/sec) at which a user can navigate to adjacent views, $w(u) = [u - \rho T, u + \rho T] = [U_L(u), U_R(u)]$ is the range of viewpoints that can potentially be displayed by the user in $n + \ell$. We call this range the *navigation window*. In order to guarantee a zero-delay view-switching, the adaptation logic at the client has to select the best set of representations such that any viewpoint in the navigation window can be reconstructed on time at the client until the next downloading request.

[1] For the sake of clarity, we formulate here the same $\mathcal{L}$ for all videos, but our work can be easily extended to the case of unequal set of representations for different videos, as shown in Section VI.B.

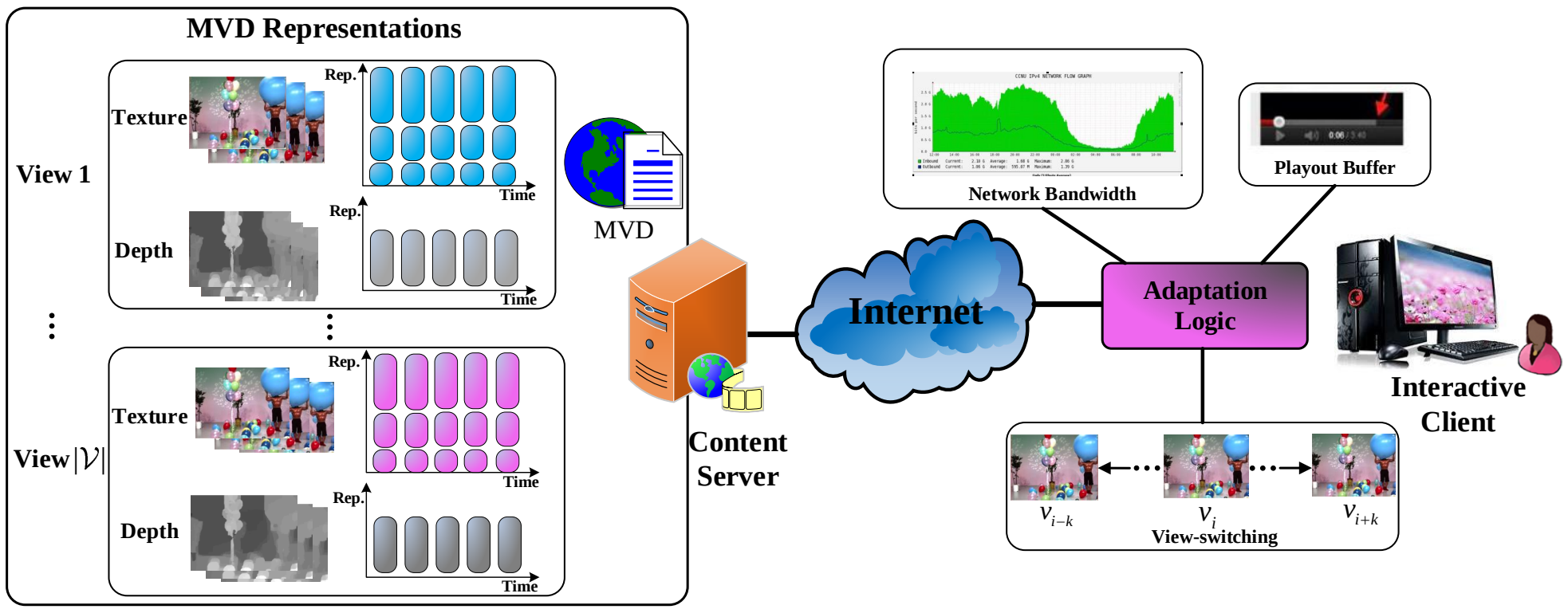

Fig. 2. Multiview-based HAS system

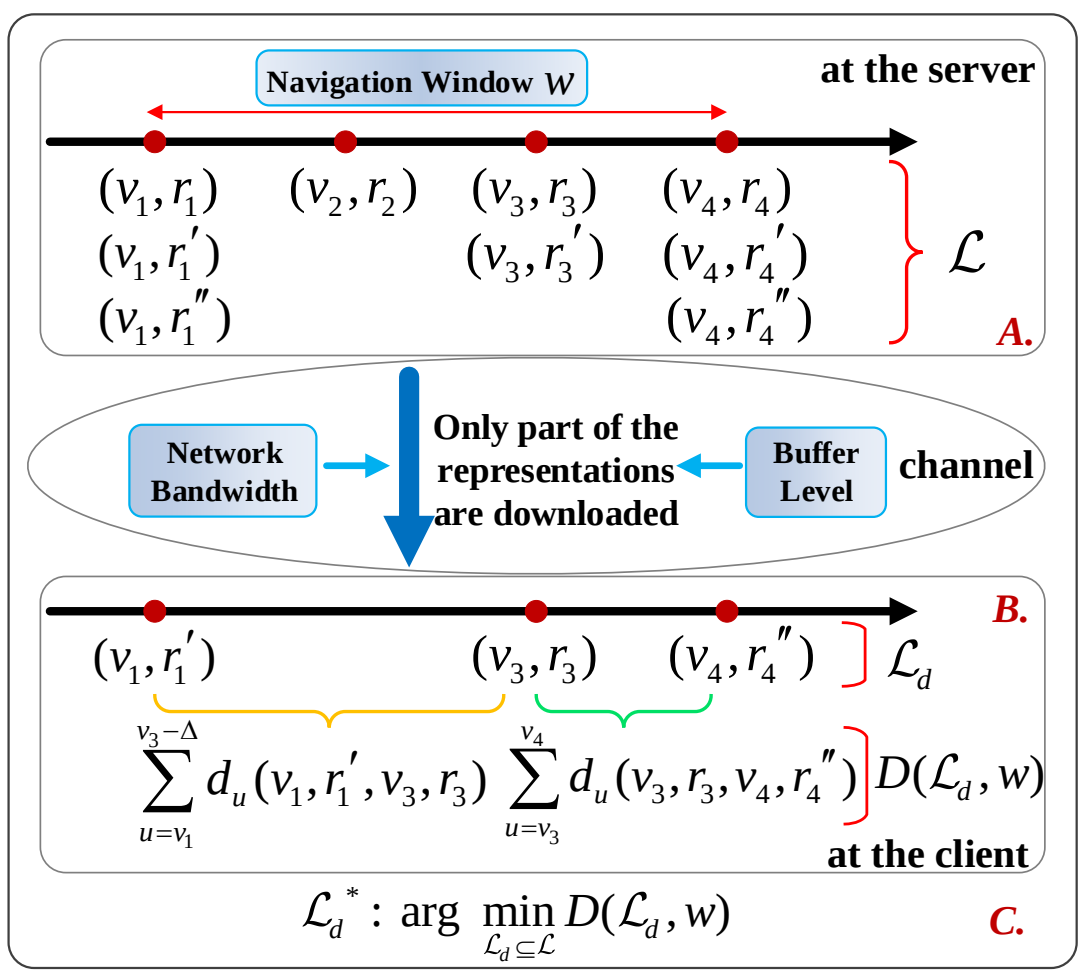

Fig. 1. Example of representation set available at the server ($\mathcal{L}$); representation set downloaded by one final user ($\mathcal{L}_d$) based on the navigation window, network bandwidth and buffer status, with the associated distortion experienced while navigating the scene; and the best representation set ($\mathcal{L}_d^*$) that minimizes the navigation distortion. A, B and C denote subsection A, subsection B and subsection C in this section, respectively.

### B. Navigation Distortion

We now evaluate the distortion experienced by a user downloading the set of representations $\mathcal{L}_d \subseteq \mathcal{L}$, while navigating in the window $w$, see part B in Fig. 1. For the sake of simplicity, in the following we do not explicitly indicate the dependency of $w$ on the current viewpoint $u$. Each viewpoint $u$ in the navigation window will be displayed at the distortion $d_u(v_i, r_i, v_j, r_j)$, where $v_i$, $v_j$ are the left and right reference views respectively, and $r_i$, $r_j$ are the corresponding encoding bitrates with $(v_i, r_i), (v_j, r_j) \in \mathcal{L}_d$. This means that a user, given $\mathcal{L}_d$, navigates in the window $w$ at the distortion

$$D(\mathcal{L}_d, w) = \frac{1}{N_u} \sum_{\substack{(v_i,v_j)\in\mathcal{L}_d \\ v_i<v_j}} \sum_{\substack{u\in w \\ u\in[v_i,v_j]}} d_u(v_i, r_i, v_j, r_j) \quad (2)$$

where $N_u$ is the number of views in the navigation window. Recall that consecutive camera views $v_i$ and $v_j$ in $\mathcal{L}_d$ are used as a pair of anchor views for all virtual viewpoints between them [33].

### C. Optimization Problem Formulation

We can now formulate a navigation-aware optimization problem for MV adaptive streaming. Given the representation set $\mathcal{L}$ available at the server, the navigation window of interest $w$ for the user and the network bandwidth constraint $C$ between the server and the user, the objective of our adaptation logic is then to optimize the set of representations $\mathcal{L}_d^*$ to download at the client, such that the navigation distortion experienced by the user is minimized, as shown in part C in Fig. 1. More formally, a particular client searches for

$$\mathcal{L}_d^* : \arg\min_{\mathcal{L}_d\subseteq\mathcal{L}} D(\mathcal{L}_d, w) \quad (3a)$$

$$\text{s.t.} \sum_{\forall i:(v_i,r_i)\in\mathcal{L}_d} r_i \leq C \quad (3b)$$

The optimal MV representations selection problem in Eq. (3) is NP-hard. This can be proven by noting that the reduced case of $|\mathcal{R}| = 1$ is shown as a camera view selection problem. This special problem can be formulated as a *set cover* (SC) problem [33], which is a NP-hard. Optimizing jointly camera view subsets and encoding bitrates is no easier than solving the SC problem, thereby the problem in Eq. (3) is also NP-hard in general cases. In the following, we propose a tractable method solving Eq. (3).

## IV. Optimal Representation Selection

To efficiently solve the optimization problem in Eq. (3), we propose an optimal solving method based on a DP algorithm with polynomial computational time complexity. Then, we provide a suboptimal greedy algorithm with effective performance to reduce even further the computational complexity.

### A. DP-based Optimization Algorithm

Given a representation $(v, r) \in \mathcal{L}_d$ and $w = [U_L, U_R]$, we define the aggregate distortion $\Phi(v, r, c)$ as the minimum distortion experienced between $\max\{U_L, v\}$ and $U_R$, when a remaining bitrate budget $c$ is available for additional reference views. Then, we can write the following iterative property:

$$\Phi(v,r,c) = \min_{(v_i,r_i), v_i>v} \left\{ \sum_{u=\max\{U_L,v\}}^{v_i-\Delta} d_u(v,r,v_i,r_i) + \Phi(v_i,r_i,c-r_i) \right\} \quad (4)$$

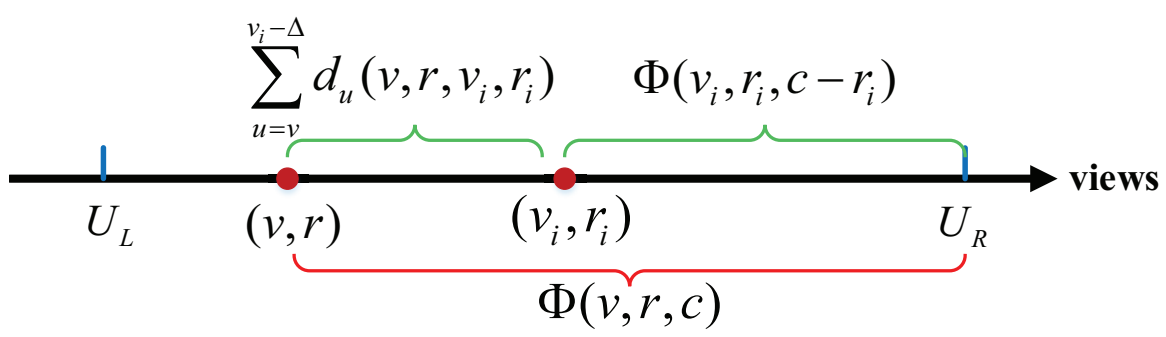

Fig. 3. The recursion property in the DP solution

The equation (4) states that, when one of the optimal representations $(v_i, r_i)$ is selected for downloading between $[v, U_R]$, the range of views $[v, U_R]$ is decomposed into two ranges $[v, v_i)$ and $[v_i, U_R]$. All viewpoints in the first range will be synthesized by the pair of camera views $v$ and $v_i$. In the second range $[v_i, U_R]$, other camera views can be selected for downloading with a total bitrate budget of $c - r_i$ as depicted in Fig. 3.

Evaluating

$$\min_{\{v_L \le U_L, r_L\}} \Phi(v_L, r_L, C - r_L) \quad (5)$$

leads to the solution of the problem in Eq. (3) under the assumptions that i) only one encoding bitrate is selected in $\mathcal{L}_d$ for each camera view, ii) the most left camera view in $\mathcal{L}_d$ is such that $v_L \le U_L$. These conditions are satisfied for most common 3D sequences [33]. Due to the recursion shown in Eq. (4), Eq. (5) can be evaluated by DP.

We can deduce the computational complexity of our solution in Eq. (4) from a bound on the size of DP table and the cost in computing each table entry. For the sake of clarity in the notation, let the number of selected reference views and the number of views covered in the navigation window be $N_v = |\mathcal{L}_d|$ and $N_u = (U_R - U_L)/\Delta + 1$, respectively. The size of the DP table $\Phi$ is no larger than $N_v \times N_c \times |\mathcal{R}|$, where $N_c$ is the number of channel bandwidth values that can be experienced during the optimization. For each entry in the DP table, we need to consider at most $(N_v - 1) \times |\mathcal{R}| + 1$ candidate camera views, and for each of them, we need to evaluate the distortion to compare. Hence the complexity in computing each entry over all navigation views is $\mathcal{O}((N_v - 1)|\mathcal{R}| + 1)N_u)$. Generally, the overall computation complexity of our proposed DP algorithm in Eq. (4) is $\mathcal{O}(N_v N_c |\mathcal{R}|((N_v - 1)|\mathcal{R}| + 1)N_u)$, which can be approximated by $\mathcal{O}(N_u N_c N_v^2 |\mathcal{R}|^2)$. Although the DP-based algorithm can achieve the optimal solution within a polynomial computational complexity, it still incurs a long execution time when the problem scale becomes large (e.g., the number of variables in $\mathcal{V}$ and $\mathcal{R}$ is large). Thus, a suboptimal solution with relatively low optimality tolerance is required in practical large scale problems.

### B. Suboptimal Algorithm

To efficiently solve the problem in Eq. (3) with lower computational complexity, we develop a suboptimal solution with effective performance. In particular, we propose an iterative algorithm, in which representations are iteratively added to the set of representations to be downloaded. The complete algorithm is provided in Algorithm 1.

**Algorithm 1** Suboptimal Algorithm

**At the beginning of each downloading opportunity:**
Given $w = [U_L, U_R]$, $\mathcal{L}$ and $C$, iteratively search for the best set $\mathcal{L}_d^{(s)} = \{(\mathcal{V}^{(s)}, \mathcal{R}^{(s)})\}$, with the best camera view subset $\mathcal{V}^{(s)}$ and encoding bitrate subset $\mathcal{R}^{(s)}$. Finally, the best subset $\mathcal{L}_d^*$ achieving the minimum distortion of the objective function in Eq. (3) is obtained.

- **Initiation:**
  Set $\mathcal{V}^{(0)} = \{v_L, v_R\}$ with $v_L \le U_L$, $v_R \ge U_R$.
- $1^{st}$ **Step:**
  Given $\mathcal{V}^{(1)} = \mathcal{V}^{(0)}$, and $s = 1$, find

$$\mathcal{R}^{(1)} = \{r_L, r_R\} : \arg\min_{\mathcal{L}_d \subseteq \mathcal{L}} D^{(1)}(\mathcal{L}_d, w) \quad (6a)$$

$$\text{with} \quad r_L + r_R \le C \quad (6b)$$

- **Search Iteration:** $(s = 2, 3, ...)$
  1) Given $\mathcal{V}^{(s-1)}$, find the least correlated view with each pair of consecutive camera views in their interval, which altogether constructs the camera views subset $\mathcal{V}^{(s)}$.
  2) Given $\mathcal{V}^{(s)}$ and $\mathcal{R}^{(s-1)}$, find the best and equal bitrate $r^{(s)}$ for all newly added camera views such that

$$\mathcal{R}^{(s)} = \hat{\mathcal{R}}^{(s-1)} \cup \{\underbrace{r^{(s)}, \cdots, r^{(s)}}_{|\mathcal{V}^{(s)}| - |\mathcal{V}^{(s-1)}|}\} : \arg\min_{\mathcal{L}_d \subseteq \mathcal{L}} D^{(s)}(\mathcal{L}_d, w) \quad (7a)$$

$$\text{with} \quad \hat{\mathcal{R}}^{(s-1)} \le \mathcal{R}^{(s-1)} - \frac{{\Delta_r}^{(s)}}{|\mathcal{V}^{(s-1)}|} \quad (7b)$$

$${\Delta_r}^{(s)} = (|\mathcal{V}^{(s)}| - |\mathcal{V}^{(s-1)}|) \cdot r^{(s)} + \sum_{\forall i: r_i \in \mathcal{R}^{(s-1)}} r_i - C. \quad (7c)$$

- **Update and Termination in each Search Iteration:**
  1) If $D^{(s)} < D^{(s-1)}$, set $s = s + 1$ and proceed to the next search iteration; otherwise, stop the iteration and output $\mathcal{L}_d^* = \mathcal{L}_d^{(s-1)}$.
  2) If $\mathcal{V}^{(s)} = \mathcal{V}$, stop the iteration and output $\mathcal{L}_d^* = \mathcal{L}_d^{(s)}$.

At each downloading opportunity, given the navigation window $w = [U_L, U_R]$, the representation set at the server $\mathcal{L}$, and the downloading bandwidth constraint $C$, a client seeks the best subset of representations to download by following a suboptimal greedy procedure. Let denote by $\mathcal{V}^{(s)}$, $\mathcal{R}^{(s)}$ and $\mathcal{L}_d^{(s)}$ respectively the camera set, the encoding bitrate set and the representation set selected at iteration $s$ of the search algorithm, with $\mathcal{L}_d^{(s)} = \{(\mathcal{V}^{(s)}, \mathcal{R}^{(s)})\}$. Starting from the lateral views $v_L$ and $v_R$ composing the best camera view subset $\mathcal{V}^{(1)}$, at step $s = 1$, the corresponding best encoding bitrates $r_L$ and $r_R$ that minimize the navigation distortion $D^{(1)}(\mathcal{L}_d, w)$ are selected by solving Eq. (6). Then, we iteratively add one camera view within each pair of anchor views. In more details, at each iteration step $s > 1$, given the best camera subset of the previous step $\mathcal{V}^{(s-1)}$, the least correlated view between each pair of previously selected consecutive camera views is selected from $\mathcal{L}$. The corresponding intermediate views are added to the previously selected views to form $\mathcal{V}^{(s)}$. For example, at step $s = 2$, the view $v_m$ ($v_L < v_m < v_R$) having the least correlation with $v_L$ and $v_R$ is added into $\mathcal{V}^{(1)}$ to compose $\mathcal{V}^{(2)} = \{v_L, v_m, v_R\}$. After finding the camera view subset $\mathcal{V}^{(s)}$, we search the best encoding bitrate $r^{(s)}$ for all newly added views based on the encoding bitrate subset $\mathcal{R}^{(s-1)}$ according to Eqs. (7a)-(7c), such that the navigation distortion $D^{(s)}(\mathcal{L}_d, w)$ is minimized. In particular, we denote by $|\mathcal{V}^{(s)}|$ the number of camera views selected at iteration $s$. Therefore, $|\mathcal{V}^{(s)}| - |\mathcal{V}^{(s-1)}|$ represents the number of newly added camera views at step $s$, and the corresponding increased bitrate is $(|\mathcal{V}^{(s)}| - |\mathcal{V}^{(s-1)}|) \cdot r^{(s)}$. On the basis of the overall bitrate $\sum_{\forall i: r_i \in \mathcal{R}^{(s-1)}} r_i$ at the previous step $s-1$, the excessive value ${\Delta_r}^{(s)}$ over the bandwidth constraint $C$ is calculated by Eq. (7c). Then, we average it over $|\mathcal{V}^{(s-1)}|$ and reduce each element in $\mathcal{R}^{(s-1)}$ by this mean value, such that the encoding bitrate set for the camera views in $\mathcal{V}^{(s-1)}$ is updated as $\hat{\mathcal{R}}^{(s-1)}$ as depicted in Eq. (7b). In such a way, the best coding subset $\mathcal{R}^{(s)}$ is finally obtained by inserting the $|\mathcal{V}^{(s)}| - |\mathcal{V}^{(s-1)}|$ encoding bitrates $r^{(s)}$ for corresponding added views into set $\hat{\mathcal{R}}^{(s-1)}$, as shown in Eq. (7a).

In summary, at each step $s$, we first look for the best camera views subset $\mathcal{V}^{(s)}$. Then the corresponding best bitrate subset $\mathcal{R}^{(s)}$ is optimized according to the objective function in Eq. (3). We define $\mathcal{L}_d^{(s)} = \{(\mathcal{V}^{(s)}, \mathcal{R}^{(s)})\}$ as the optimal representation set at step $s$ and the resulting distortion is $D^{(s)}$. This procedure is repeated iteratively until the navigation distortion no longer reduces ($D^{(s)} > D^{(s-1)}$) or no more view can be added ($\mathcal{V}^{(s)} = \mathcal{V}$). Finally, the best representation set $\mathcal{L}_d^*$ able to minimize the navigation distortion is obtained. In fact, to reduce the computational complexity, we allocate an equal bitrate for all added camera views at each step, which possibly is not the optimal bitrate. For this reason, our greedy algorithm is necessarily suboptimal.

The computational complexity of the proposed suboptimal algorithm is no larger than $\mathcal{O}((S+1)|\mathcal{R}|)$, where $S$ is the number of iterations. As the value of $S$ increases, the complexity of the suboptimal algorithm becomes larger but the performance generally improves. This strictly depends on the video content characteristics, as a video highly affected by virtual synthesis artifacts ideally requires more iterative steps. In addition, the value of $S$ probably increase with the size of the representation set and navigation window. It is worth noting that, even in the worst case, the number of iterations $S$ is upper bounded by only $\lceil (U_R - U_L)/2\Delta \rceil + 1$. This therefore indicates a substantial complexity reduction compared to the DP-based optimization algorithm in Section IV.A. As we will show in the simulation results in Section VI, this algorithm yet offers good performance in practice.

## V. Adaptive Streaming Framework

### A. Simulated Settings

Our simulations are carried out with three multiview video sequences at $1080p$ resolution, namely "*Dancer*" [34], "*Shark*" [35], and "*Hall*" [36]. The sequences are highly heterogenous in terms of coding and view synthesis efficiency, and are thus representative of various video categories. For example, "*Dancer*" is a very dynamic sequence highly affected by coding artifacts, while "*Hall*" is a quite static scene but with a 3D geometry that makes virtual view synthesis highly challenging. For each video sequence, we consider two types of representation sets that can be stored at the server, defined as "Ad-hoc Set" and "Optimized Set". In particular, for the "Ad-hoc Set", we assume two sets of representations, namely $\mathcal{L}1$ and $\mathcal{L}2$ provided in Table II, with $\mathcal{L}1$ being a larger set (i.e., more representations per video) than $\mathcal{L}2$ and having therefore a greater search space. For the "Optimized Set", the representation set is optimized a priori at the server given storage constraints and clients population as in [22]. The solver IBM ILOG CPLEX [37] is used to obtain the optimal solution, with unequal optimized sets being stored for different videos.

TABLE II
The "Ad-hoc Set" available at the server

| | $\mathcal{L}1$ | $\mathcal{L}2$ |
|---|---|---|
| Camera View ID | 1 2 3 4 5 6 7 8 9 10 | 1 3 5 7 10 |
| Encoding Bitrate (Mbps) | 0.1 0.2 0.3 0.5 1 2 3 4 6 8 10 12 15 18 20 | 0.1 0.3 1 3 6 10 15 |

The navigation scenarios considered in our simulations are characterized by *static* or *dynamic* navigation. The former means that the navigation window is constant over the entire streaming session, while the latter considers a dynamic behavior of navigation windows. In this case, the navigation path evolves over time following a dynamic navigation model in [38] with the minimum space between two adjacent views $\Delta = 0.1$. More specifically, we simulate i) a *uniform* navigation, where the user has the same probability of remaining the current view, or switching to the left or right view, and ii) a *non-uniform* navigation, where the user has a probability $p_n$ of remaining the current view and $(1-p_n)/2$ of switching to the left or right view.

We finally consider the *dynamic* network bandwidth, which varies over time following either a Markovian model [39], [40] or a channel prediction model. The bandwidth set is $\{0.6, 1, 2, 3, 4, 5, 6, 8, 10\}$(Mbps) in the former. We set the Markov transition matrix that allows transitions to adjacent states with probability $2p_c/3$ and two-state jumps with probability $p_c/3$.In the latter, we consider a two-stage exponentially weighted moving average (EWMA) predictor [41]. Based on the channel information of previous video segments, the drift $\Delta\hat{C}(n)$ and absolute value $\hat{C}'(n)$ of bandwidth for the current segment $n$ are estimated separately by Eqs. (8)-(10) in Algorithm 2. Then, the final estimation of the bandwidth $\hat{C}(n)$ is obtained by combining two estimated values as shown in Eq. (11). In particular, $\tilde{C}(n)$ is the realistic TCP throughput, which can be calculated by $\tilde{C}(n) = \sum_i r_i(n) \cdot \tau / \tilde{T}(n)$, where $\tilde{T}(n)$ is the duration for downloading the sum of bitrate $r_i(n)$ for the video segment $n$.

We categorize the simulated scenarios based on the interactivity level of the user as follow:

- "Low-interactivity Scenario": users have a *static* navigation window, e.g., $w = [5.5, 6.5]$ or $w = [1.5, 9.5]$, but

TABLE III
FITTING PARAMETERS FOR INDEPENDENT CODING IN EQ. (14)

| Video | a | b | e |
|---|---|---|---|
| *Shark* | 1 | 745.90 | 1192.10 |
| *Dancer* | 0.98 | 282.17 | 469.13 |
| *Hall* | 0.98 | 129.89 | 544.39 |

experience the *dynamic* network bandwidth. It allows us to explicitly compare the difference of optimal representation sets for each video sequence selected by different algorithms.

- "High-interactivity Scenario": users have *dynamic* experiences in terms of both navigation paths and channel bandwidth. This scenario aims to analyze the influence of different interactivity levels in the representation selection optimization.

In both scenarios, for each realization of the channel and user navigation path, the adaptation logic is run at each downloading opportunity by the client.

We consider both of above scenarios in theoretical and in more realistic settings. Simulation in theory aims to definitely analyze the crucial strategy that differentiates our optimization from the baseline ones in a stationary regime. We then also evaluate the utility that our adaptation logic offers in realistic scenarios, which incorporate realistic network statistics, buffer levels and the server-side optimization.

- In theoretical settings: we test the proposed adaptation logic with the "Ad-hoc Set". $|\mathcal{N}| = 50$ video segments ($50\tau$ seconds) are considered for each video sequence. Dynamic bandwidth varies over time following the Markovian model. We assume infinite playback buffers and exact channel estimation.
- In realistic settings: both "Ad-hoc Set" and "Optimized Set" are simulated. We consider the datasets from Neubot data received by the Measurement Lab [42] as the realistic channel traces in our simulation. To this end, the channel prediction model is employed to estimate the network bandwidth. Algorithm 2 shows the adaptation logic in this realistic setting. At the beginning of each downloading opportunity $n$, the available bandwidth is first estimated using the two-stage EWMA predictor. Then, the buffer is taken into account in deciding when to schedule the next downloading time from Eq. (12). In particular, $\hat{T}(n)$ is the target inter-request time. $B_0$ is the buffer reference level, towards which the buffer level attempts to converge to (we set as 20s), see [43] in details. Finally, given the estimated bandwidth, we can proceed to optimize Eq. (3).

### B. Synthesis Distortion Metrics

To evaluate the navigation distortion $D(\mathcal{L}_d, w)$ given by Eq. (2), we adopt the synthesis distortion model from [22], provided in the following:

$$d_u(v_i, r_i, v_j, r_j) = \alpha D_{\min} + (1-\alpha)\beta D_{\max} + [1-\alpha-(1-\alpha)\beta] D_I \quad (13)$$

where $D_{\min} = \min\{D(v_i), D(v_j)\}$, $D_{\max} = \max\{D(v_i), D(v_j)\}$, $D_I$ is the inpainted distortion, and $D(v_i), D(v_j)$ are the distortion of left and right reference views for the synthetic view $u$, respectively. Here, $\alpha = \exp(-\xi|u - v_{\min}|)$, and $\beta = \exp(-\xi|u - v_{\max}|)$, with $v_{\min} = v_i$, $v_{\max} = v_j$ if $D(v_i) \leq D(v_j)$, otherwise, $v_{\min} = v_j$, $v_{\max} = v_i$ if $D(v_i) > D(v_j)$. The parameters $\xi$ and $D_I$ can be evaluated by curve fitting. Similarly, we set the inpainting distortion to $D_I = 0.35$, and $\xi = \{0.35, 0.52, 1.32\}$ for "*Dancer*" ("sport-action" video), "*Shark*" ("cartoon" video), and "*Hall*" ("movie" video), respectively.

Finally, the distortion of the coded camera view $v_i$ encoded at bitrate $r_i$ follows the model

$$D(v_i) = 1 - (a - \frac{b}{r_i + e}) \quad (14)$$

where $a$, $b$ and $e$ are parameters that depend on both the content characteristics and the resolution of the video, as shown in Table III. These are set to fit experimental $(1-\text{VQM})$ data points, where VQM is an objective Video Quality Metric [22]. Note that a visually pleasant video usually has a VQM score below 0.2 and a reduction in VQM of 0.1 is a significant quality improvement as shown in [44].

---

**Algorithm 2** Adaptation Logic in more realistic settings

**At the beginning of each downloading opportunity $n$:**

- Estimate the bandwidth $\hat{C}(n)$ by solving:

$$\Delta\tilde{C}(n-1) = \tilde{C}(n-1) - \tilde{C}(n-2). \quad (8)$$

$$\Delta\hat{C}(n) = (1-\alpha_{TCP})\Delta\hat{C}(n-1) + \alpha_{TCP}\Delta\tilde{C}(n-1). \quad (9)$$

$$\hat{C}'(n) = (1-\beta_{TCP})\hat{C}(n-1) + \beta_{TCP}\tilde{C}(n-1). \quad (10)$$

$$\hat{C}(n) = \hat{C}'(n) + \Delta\hat{C}(n). \quad (11)$$

- Determine the target time until the next request $\hat{T}(n)$ by:

$$\hat{T}(n) = \frac{\sum_i r_i(n) \cdot \tau}{\hat{C}(n)} + \kappa(B(n-1) - B_0). \quad (12)$$

- Optimize Eq. (3) given $\hat{C}(n)$.

---

### C. Baseline Algorithms

Our proposed algorithm is compared to three adaptation logics proposed in the literature, namely, in [30] labeled as "view adaptation", in [8] labeled as "rate adaptation" and an algorithm extrapolated by [8] labeled as "2views-based rate adaptation". The "view adaptation" method encodes representations with the 3D-HEVC format and then selects the best subset of camera views under a total channel constraint. This method is efficient in the coding process (due to the joint coding), but it imposes the equal encoding bitrate for all selected camera views. Moreover, each subset of jointly encoded representations (provided by the server) targets to cover the entire set of camera views, whereas the navigation window in our case changes over time and is not forced to cover the entire set. Therefore, this adaptation cannot be directly implemented and extended to our case. To this end, we propose a similar version. We subdivide the entire camera set into segments by two-camera size, i.e., each segment includes two camera views. Cameras in the segments are jointly encoded at different total bitrate values using constant bit rate (CBR), where all views in each segment have an equal bitrate following the presentation sets in Table II. The fitting parameters in Eq. (14) for joint coding process are shown in Table IV. Then at each downloading opportunity, the best subset of segments covering the navigation window are requested. In the "rate adaptation" algorithm, a two-step algorithm is used. In the first step, the set of camera views is selected, where only a pair of views are downloaded. A third camera view is also prefetched eventually in case we predict that the navigation will go outside the range covered by the first two views. In the second step, the bitrate per camera view is optimized. We assume that all viewpoints in the navigation window that cannot be covered by the two or three downloaded cameras are synthesized using only one of lateral views [45], i.e., $D_{\max} = 0$ in Eq. (13). Furthermore, we extend this algorithm to the case of navigation window and label the resulting method as the "2views-based rate adaptation". This means that we first select lateral views that better cover the navigation window for one video segment, then the corresponding encoding bitrates are optimized to minimize the navigation distortion subject to the channel constraints.

TABLE IV
FITTING PARAMETERS FOR JOINT CODING IN EQ. (14)

| | $\mathcal{L}1$ | | | $\mathcal{L}2$ | | |
|---|---|---|---|---|---|---|
| Video | a | b | e | a | b | e |
| *Shark* | 1 | 544.78 | 891.90 | 1 | 614.70 | 1073.1 |
| *Dancer* | 0.99 | 301.47 | 662.24 | 0.98 | 263.23 | 498.45 |
| *Hall* | 0.99 | 160.01 | 843.10 | 0.99 | 147.30 | 633.67 |

## VI. SIMULATION RESULTS

We now study the performance of our proposed solutions in the scenarios presented in Section V.A, and evaluate their performance in terms of VQM distortion metric [22].

### A. Results in theoretical settings

We first compare the proposed adaptation logic with the baseline methods in the case of "Low-interactivity Scenario". This particular scenario is the most favorable one for the competitor algorithms, which do not fully take into account interactivity in their optimization. In Fig. 4, the expected distortion as a function of the available bandwidth is provided for navigation window (a) $w = [5.5, 6.5]$ and (b) $w = [1.5, 9.5]$ when the representation set $\mathcal{L}1$ is available at the server. Simulation results are provided for our optimal algorithm (solid lines) as well as competitor algorithms, i.e., "view adaptation" (dotted lines) and "2views-based rate adaptation" (broken lines). It can be observed that, even in this particularly static scenario, the proposed optimal algorithm always outperforms the baseline ones for any channel constraint with a gain up to 0.13 with respect to "view adaptation" algorithm for "*Shark*" and a gain up to 0.1 with respect to "2views-based rate adaptation" for "*Hall*" at $w = [5.5, 6.5]$. While for $w = [1.5, 9.5]$, it has a gain up to 0.06 and 0.18, respectively. This is because our method is able to find the right tradeoff between coding and synthesis artifacts. Note that at $w = [1.5, 9.5]$, when the available bandwidth is low (e.g., less than 4Mbps), the "view adaptation" outperforms the proposed one for "*Hall*". The reason is that, at low bandwidth, most of the representations have the same (low) encoding bitrate. Therefore, the "view adaptation" method, which imposes equal bitrate across views, does not largely penalize the performance. It actually gains in terms of coding efficiency due to the joint encoding. However, as the bandwidth increases, the joint coding algorithm with the equal bitrate among views cannot offer a better usage of the bandwidth, and then our optimal method with adaptive bitrates achieves a better quality. Furthermore, this "view adaptation" algorithm is not efficient neither in the case of $w = [5.5, 6.5]$. Because it has to simultaneously select the camera view 8 when selecting view 7 into the optimal representation set, due to the joint encoding between them, but view 8 actually is of no help. Therefore this approach does not provide much flexibility in terms of both rate adaptation and view adaptation. For the details on quality comparison in terms of PSNR and SSIM, please refer to Appendix A.

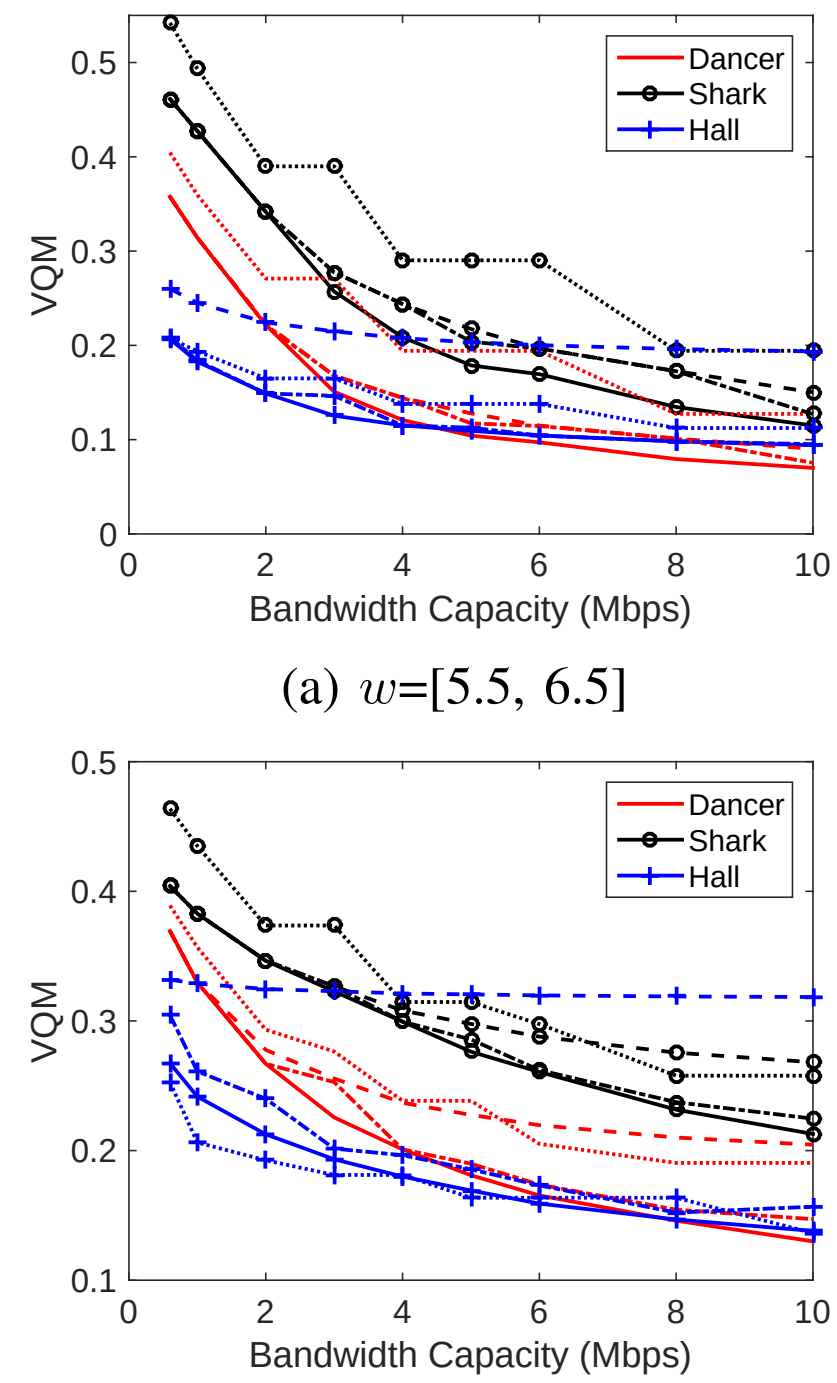

(a) $w$=[5.5, 6.5]

(b) $w$=[1.5, 9.5]

Fig. 4. Distortion comparison of a client having navigation window (a) $w = [5.5, 6.5]$ and (b) $w = [1.5, 9.5]$ vs. bandwidth capacities using $\mathcal{L}1$ in the case of "Low-interactivity Scenario". Solid lines show the performance of our optimal algorithm, while dotted lines, broken lines and dotted plus broken lines show the performance of "view adaptation", "2views-based rate adaptation" and our suboptimal algorithm, respectively.

To better understand this tradeoff, in Fig. 5 we provide the optimal representation sets for each video sequence selected by different algorithms, when the channel constraint is set to $C$ =10Mbps. Each point along the curves is an additional representation whose camera view index is indicated in the x-axis and its encoding bitrate is indicated in the y-axis. For the "*Dancer*" sequence, the proposed optimization selects five views at medium-high bitrates to cover the navigation window, while a larger number of views at lower bitrates are selected for the "*Hall*" sequence. This is due to the fact that "*Dancer*" is highly affected by coding artifacts (due to the high-motion content) and not drastically by the synthesis artifacts (due to a simple scene geometry). On the contrary, "*Hall*" has the largest dissimilarity among adjacent camera views, making the synthesis process highly challenging. Therefore, many camera views are selected in such a way that virtual viewpoints are always synthesized by close-by anchor views. To meet the channel constraints, the camera views downloaded for "*Hall*" are the ones encoded at lower bitrate. Therefore, the joint optimization of both camera views and encoding bitrates leads to an unequal allocation of the 10Mbps available per sequence, based on the content characteristics. This unequal allocation is exactly what differentiate our strategy from the baseline ones. The "view adaptation" method is limited to the same bitrate for all the views and most of the time selects many views but at low encoding bitrate. This might be convenient for the "*Hall*" sequence but not for the other ones. On the contrary, the "2views-based rate adaptation" leads to a limited number of downloaded views but at high encoding bitrate. This can be suitable to optimal camera selection for "*Dancer*" but not for the "*Hall*" sequence.

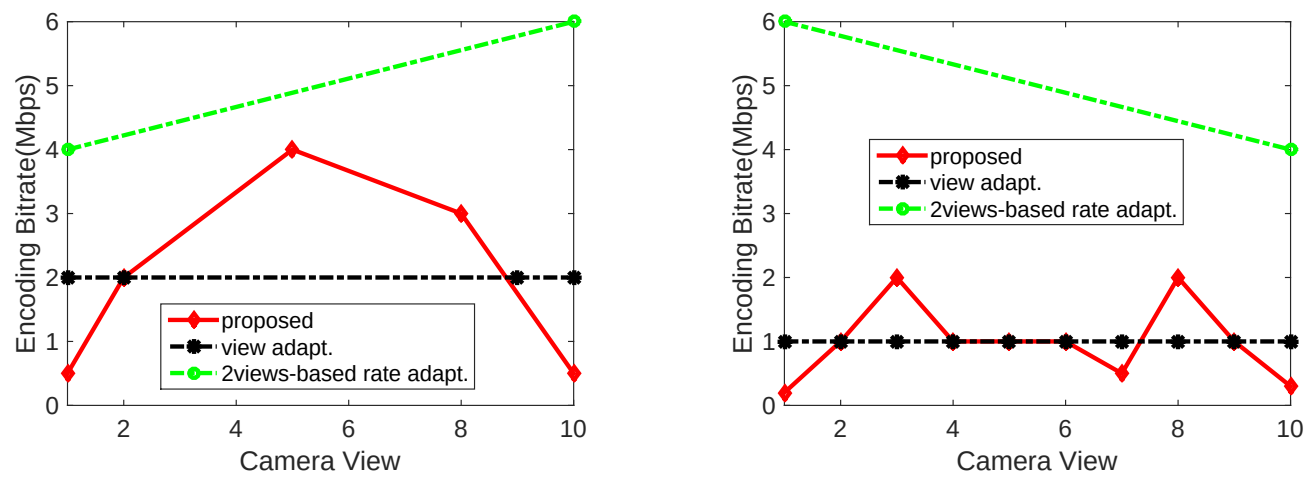

(a) *Dancer* (b) *Hall*

Fig. 5. Comparison of selected optimal representation sets with navigation window $w = [1.5, 9.5]$ at $C$ =10Mbps using $\mathcal{L}1$.

We now consider more interactive scenarios, i.e., "High-interactivity Scenario", where users navigate within the 3D scene. This leads to a variation of the navigation window over time. We simulate three types of navigation paths: (1) a *uniform* navigation with view 2.4 as first viewpoint; (2) a *non-uniform* navigation with $p_n = 0.3$ and view 2.4 as starting point; (3) a *non-uniform* navigation with $p_n = 0.6$ and view 5.1 as initial view. It means that the third one is in favor of the bigger window, since the initial view is close to the center of camera sets stored at the server in Table II. To better understand the effect of dynamic paths on quality variations, we first depict the distortion experienced over time for a bandwidth constraint realization. This dynamic bandwidth is randomly generated by the channel model with $p_c = 0.5$ and indicated in the right y-axis of Fig. 6. For each video segment downloaded progressively, the experienced distortion for the navigation window of interest is provided in the left y-axis of Fig. 6 for the "*Hall*" video sequence and with $\mathcal{L}1$ as the set of representations available at the server. As expected, the greater the available bandwidth the lower the distortion. Most importantly, despite the low channel bandwidth values, our adaptation logic outperforms the baseline ones since it is able to adapt its requests to the interactivity of the users. Gains up to 0.05, 0.08 and 0.19 are achieved in the second *non-uniform* navigation case with respect to "view adaptation", "rate adaptation" and "2views-based rate adaptation" logic, respectively. Note that, there is a sudden quality degradation in "rate adaptation" algorithm in the case of highly active users, due to the synthesis using only one of references for the navigation viewpoint out of the covered range.

To conclude, we test our proposed adaptation logic for different "Ad-hoc Set" stored at the server as well as d-ifferent video sequences. Dynamic channels are considered with $p_c = \{0.25, 0.5, 0.75, 0.9\}$. We average the distortion over 10000 realizations (100 navigation runs $\times$ 100 channel runs) to get statistically meaningful results. For the sake of clarity, we only provide the results of a navigation case per sequence in Fig. 7, where the above and below ones are for $\mathcal{L}1$ and $\mathcal{L}2$ respectively. In both scenarios, the performance of the proposed adaptation algorithm substantially outperforms that of all comparative algorithms for all categories of clients. The distortion becomes larger in the case of a more limited representation set (see the below ones). This is expected since the small set $\mathcal{L}2$ reduces the search space in the optimization as well as the room for finding optimal solutions. However, when using $\mathcal{L}1$, the overall mean VQM distortion reduction that we achieve is up to 0.06 with respect to the "view adaptation" algorithm for "*Shark*", 0.03 and 0.13 with respect to the "rate adaptation" and "2views-based rate adaptation" logic for "*Hall*", respectively. Meanwhile, we can also achieve the distortion reduction up to 0.1, 0.04 and 0.14 respectively using $\mathcal{L}2$. We recall that a distortion reduction of 0.1 in terms of VQM points is considered to be a significant improvement.

### B. *Results in realistic settings*

To evaluate the performance of the proposed suboptimal algorithm, we first provide the resulting distortion in the case of "Low-interactivity Scenario" in Fig. 4. It can be observed that, our suboptimal algorithm (dotted plus broken lines) can achieve a very close performance with respect to the optimal algorithm and outperforms the baseline ones in both navigation window cases. Except for "*Hall*" at bigger window $w = [1.5, 9.5]$, the "view adaptation" method shows better performance when the bandwidth constraint is strict (e.g. smaller than 6Mbps). We recall the fact that joint coding is more efficient for "*Hall*" having highly challenging synthesis process in low-bandwidth case, as the search space as well as the room for finding optimal solutions is extremely limited.

We now pick realistic network statistics from Neubot data as a specific channel trace in Fig. 8 to better show the temporal evolution of the users satisfaction. The two-step EWMA predictor is used to estimate the available bandwidth at the beginning of each downloading opportunity, and we show its performance of prediction in Fig. 8. In Fig. 9, we compare the simulated VQM distortion over time using $\mathcal{L}1$ in the second case of the *non-uniform* navigation model. It is again verified that, for three sequences, our optimal logic substantially outperforms all baselines and achieves the best navigation quality over time. In particular, for "*Dancer*", gains up to 0.04, 0.07 and 0.07 in terms of average distortion over time are achieved with respect to "view adaptation", "rate adaptation" and "2views-based rate adaptation" logic, respectively. At the same time, the proposed suboptimal algorithm outperforms three competitor algorithms in general with gains up to 0.02, 0.05 and 0.06 for "*Dancer*". For "*Hall*", We notice that the "view adaptation" method with joint coding offers the less distortion in few temporal instants characterized by low estimated bandwidth (e.g., from 20s to 40s). However, a growing loss at the quality level is experienced when the bandwidth increases. Compared to the "rate adaptation" and "2views-based rate adaptation" logic, the suboptimal algorithm can still yield the gains up to 0.11 and 0.15, respectively. This is because our suboptimal solution carefully considers both coding and view synthesis artifacts. The results further considering the quality variations are provided in Appendix B. At the same time, in Appendix C, we show that our video player maintains reasonable buffer occupancy without risking a buffer underflow and overflow.

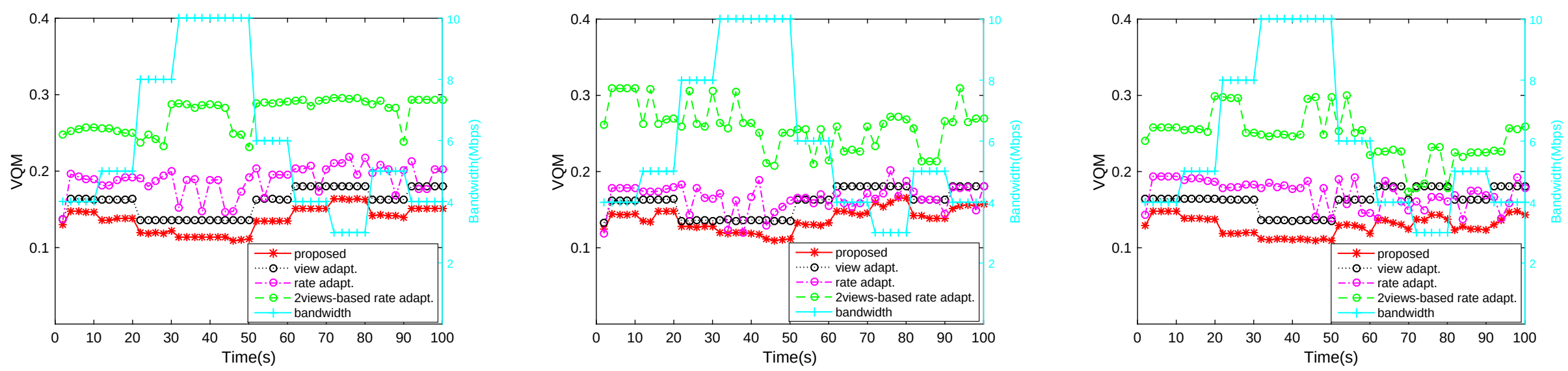

(a) *uniform* (b) *non-uniform* with $p_n = 0.3$ (c) *non-uniform* with $p_n = 0.6$

Fig. 6. Distortion comparison over time with a specific channel realization using $\mathcal{L}1$ for "*Hall*" in the case of "High-interactivity Scenario".

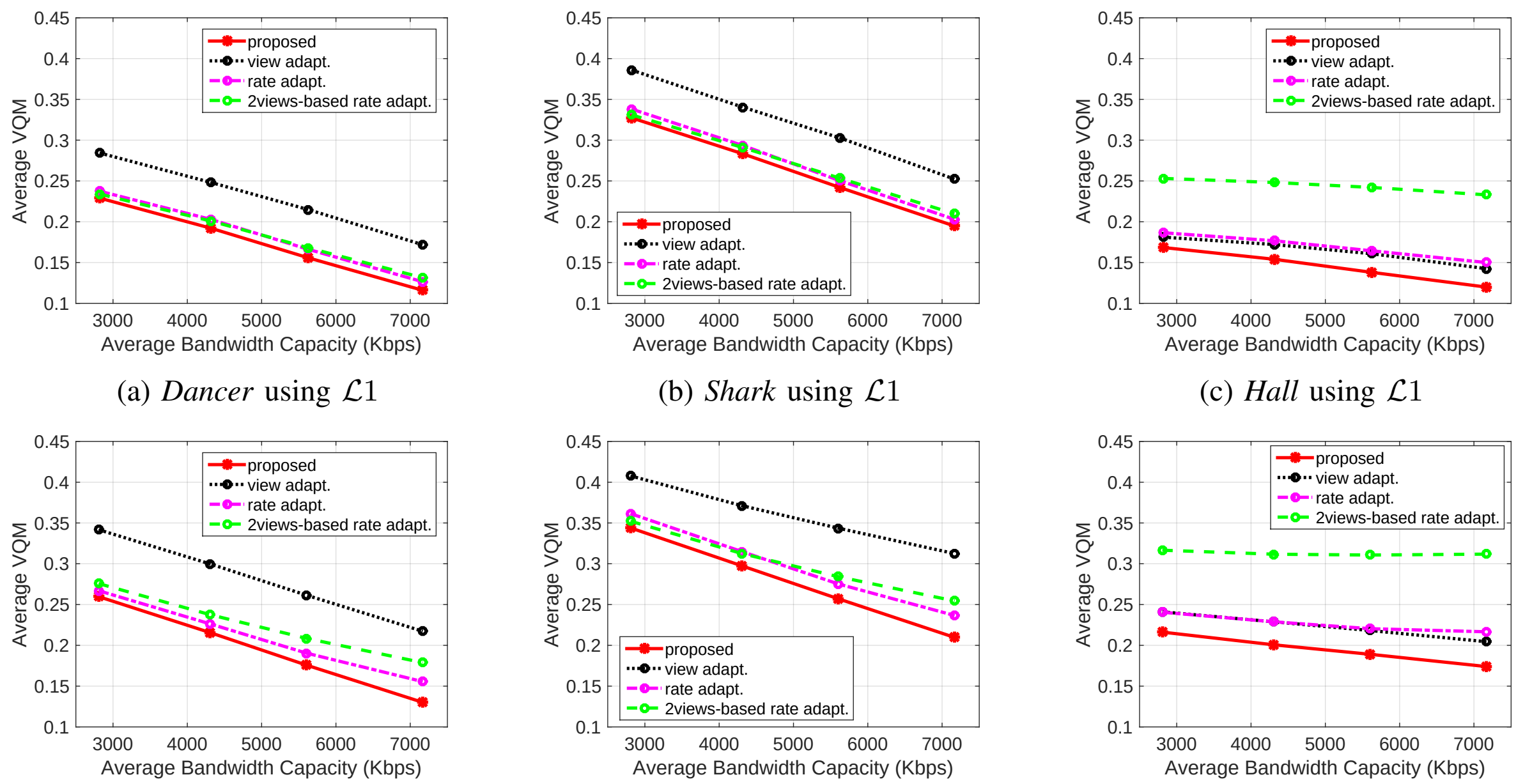

(a) *Dancer* using $\mathcal{L}1$ (b) *Shark* using $\mathcal{L}1$ (c) *Hall* using $\mathcal{L}1$

(d) *Dancer* using $\mathcal{L}2$ (e) *Shark* using $\mathcal{L}2$ (f) *Hall* using $\mathcal{L}2$

Fig. 7. Comparison of average distortion over time for the "High-interactivity Scenario". (a) and (d) *Dancer* in the case of *uniform*, (b) and (e) *Shark* in the case of *non-uniform* with $p_n = 0.3$, as well as (c) and (f) *Hall* in the case of *non-uniform* with $p_n = 0.6$.

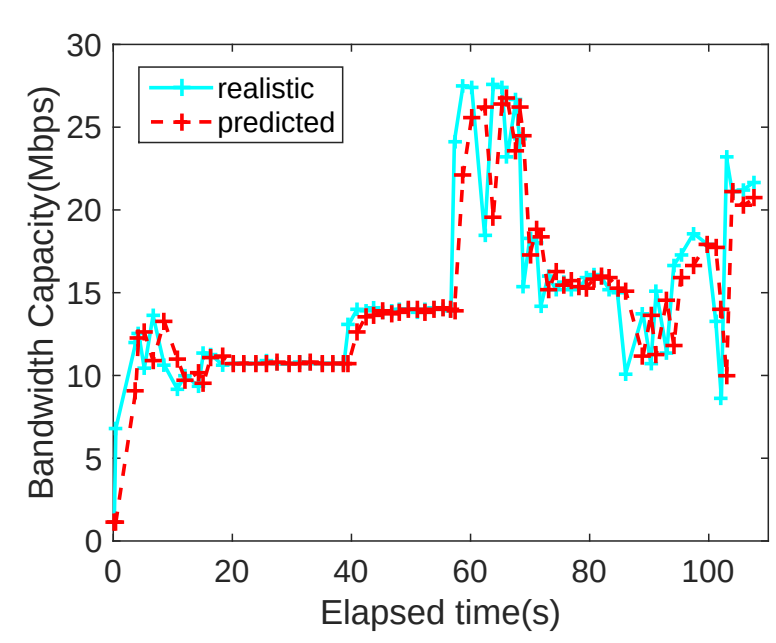

Fig. 8. The realistic bandwidth over time in a specific channel trace and predicted bandwidth by the two-step EWMA.

To see the visual results, we pick the decoded frame of synthesized view $u = 4.5$ of "Dancer" at elapsed time $t = 5.5$s with navigation window $w = [1.5, 8.7]$ and the bandwidth capacity $C = 12$Mbps. Then we show the enlarged portions In Fig. 10. It can be seen that, the closest visual results to the original image are given by our proposed algorithms, especially by comparing the regions in the red rectangles. The VQM distortion is 0.09 and 0.1 obtained by our proposed optimal and suboptimal algorithm, respectively. The "view adaptation" method provides the lower quality with VQM = 0.15, since multiple views encoded at a low encoding bitrate are selected. Due to the rendering using a single reference view for the navigation viewpoint out of the covered range, the "rate adaptation" algorithm leads to a considerable amount of holes in the virtual view with VQM = 0.13, particularly in the boundaries. Finally, in the "2views-based rate adaptation" method, the selected lateral views of the navigation window are basically difficult to reconstruct the middle view at a high quality and edge artifacts are obviously introduced, resulting in VQM = 0.22.

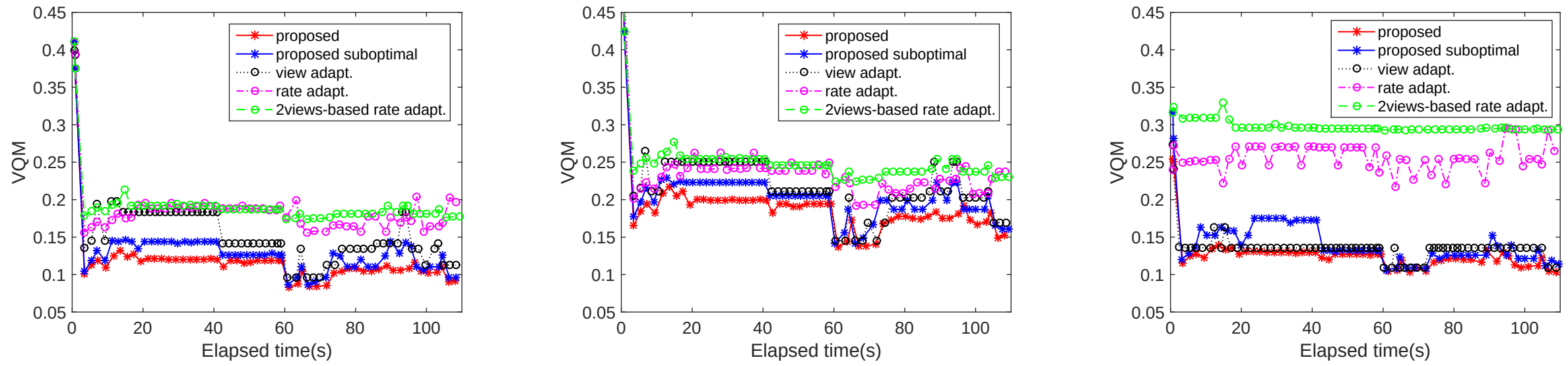

(a) *Dancer* (b) *Shark* (c) *Hall*

Fig. 9. Distortion comparison over time in the case of *non-uniform* navigation model with $p_n = 0.6$ and a specific channel trace in Fig. 8 using $\mathcal{L}1$ for the "High-interactivity Scenario".

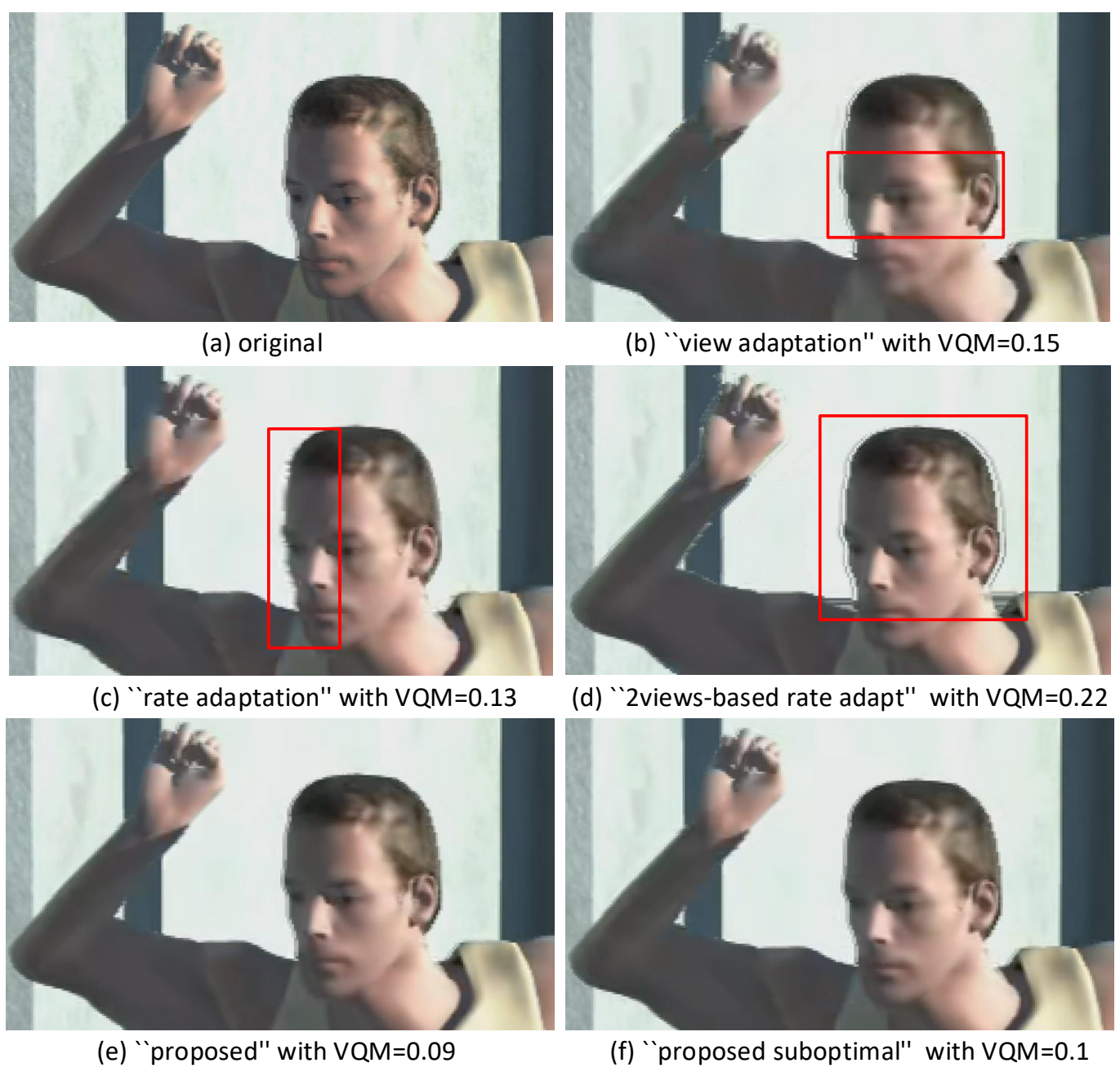

Fig. 10. Visual comparison of the enlarged portions in the decoded frame of synthesized view $u$ = 4.5 of "Dancer" at elapsed time $t$ = 5.5s. From the regions in the red rectangles, we can observe the blurring effect given by (b) the "view adaptation" method, a considerable amount of holes provided by (c) the "rate adaptation" algorithm, and edge artifacts generates by (d) the "2views-based rate adaptation" method.

To overall gain insight into the difference among the algorithms, we provide the results in terms of average VQM distortion for different "Ad-hoc Set" under five channel traces. Each simulation for a navigation model is averaged over 100 runs under a channel trace. The results are shown in Fig.11, where the above and below ones are respectively for $\mathcal{L}1$ and $\mathcal{L}2$. Compared to all comparative algorithms, the proposed optimal solution again achieves substantial gains for all categories of clients in both scenarios. When using $\mathcal{L}1$, the overall mean distortion reduction that we achieve is up to 0.07 compared with the "view adaptation" method for "*Shark*", 0.11 and 0.19 for "*Hall*" with respect to the "rate adaptation" and "2views-based rate adaptation" logic, respectively. Meanwhile, we can also achieve the distortion reduction of up to 0.08, 0.08 and 0.14 respectively using $\mathcal{L}2$. Most importantly, we observe that the performance of our suboptimal algorithm is very close to the optimal performance. It generally outperforms all baseline algorithms in all cases except at very low bandwidth. The distortion reduction is up to 0.06, 0.10 and 0.18 respectively when using $\mathcal{L}1$. There is a more limited gain obtained by using $\mathcal{L}2$ with more constrained representations. However, we can also achieve gains up to 0.06, 0.08 and 0.13 respectively. Similarly, the benefits of our suboptimal solution are limited for "*Hall*" when compared to the joint coding adaptation logic for both sets. Note also that the first channel trace (i.e., the average bandwidth constraint is 3.752Mbps) using $\mathcal{L}2$ is an extremely challenging scenario, where limited encoding bitrates are provided in the selection. For this reason, our suboptimal logic has the similar mean VQM with respect to the "rate adaptation" scheme for "*Dancer*" and "*Shark*", which are highly affected by coding artifacts.

After illustrating the gains captured by our adaptation logic in terms of the "Ad-hoc Set", we now provide results in the case of the representation set optimized a priori at the server, i.e., the "Optimized Set". We first consider the "Low-interactivity Scenario", with navigation window $w = [1.5, 9.5]$, five channel traces in Fig. 11 being considered, and the storage constraint of the server per video being 18Mbps. The representation optimization for this scenario is then performed at the server side and described in Appendix D. Given the diverse "Optimized Set" for different videos, where unequal encoding bitrates are allocated for different views as well, we compare the performance of different adaptation logics. The average VQM distortion for users experiencing different channel traces is shown in Fig. 12. It can be seen that, both the proposed optimal and suboptimal algorithm sig-nificantly outperform the baseline for any channel constraint with gains up to 0.2 and 0.19 for "*Hall*" respectively.

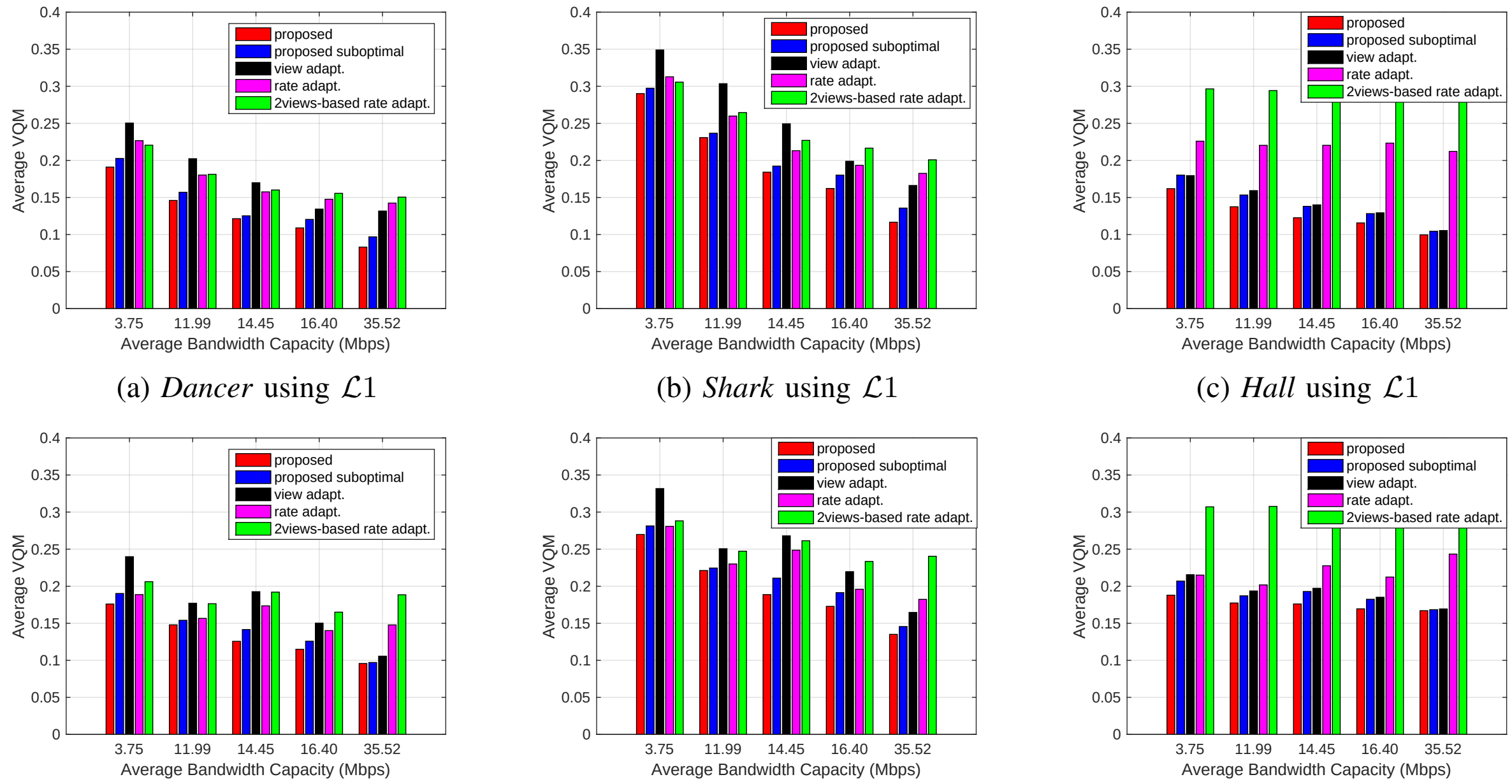

(a) *Dancer* using $\mathcal{L}1$ (b) *Shark* using $\mathcal{L}1$ (c) *Hall* using $\mathcal{L}1$

(d) *Dancer* using $\mathcal{L}2$ (e) *Shark* using $\mathcal{L}2$ (f) *Hall* using $\mathcal{L}2$

Fig. 11. Comparison of average distortion over time for the "High-interactivity Scenario". (a) and (d) *Dancer* in the case of *uniform*, (b) and (e) *Shark* in the case of *non-uniform* with $p_n = 0.3$, as well as (c) and (f) *Hall* in the case of *non-uniform* with $p_n = 0.6$.

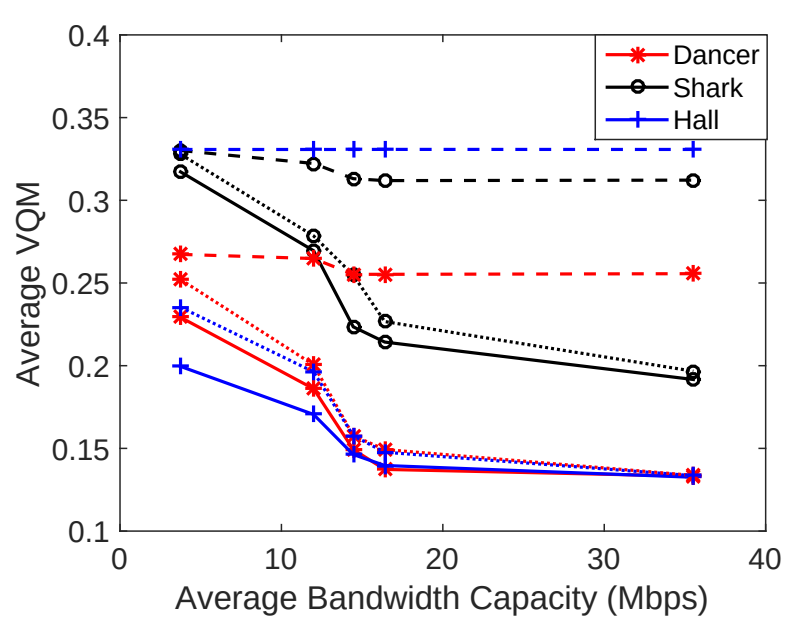

Fig. 12. Distortion comparison of different algorithms based on the "Optimized Set" for the "Low-interactivity Scenario" with $w = [1.5, 9.5]$. Solid lines show the performance of our optimal algorithm, while dotted lines, broken lines show the performance of the suboptimal adaptation and "2views-based rate adaptation"logic, respectively.

We now consider the "High-interactivity Scenario". The representations are first optimized at the server with dynamic navigation windows following the model in [22]. The storage constraint per video is 18Mbps as well. For the details, please refer to Appendix D. At the client side, based on the "Optimized Set" of this scenario, 5 types of users having variable navigation windows and different channel traces are considered for each video and navigation model. For each type of user, 100 users are simulated and averaged to obtain the mean VQM distortion. The distortion comparison of different algorithms in one navigation case per sequence is depicted in Fig. 13. It can be shown that, both the proposed optimal and suboptimal algorithm substantially outperform two comparative algorithms for all categories of clients. Note that, "*Dancer*" and "*Shark*" have large VQM distortion in "rate adaptation" method (pink bar). This is because these two sequences highly suffer from coding artifacts, resulting in a limited number of views but at higher encoding bitrates stored in the "Optimized Set" via the server-side optimization. Given the bandwidth constraints, the "rate adaptation" method thus probably would not find the proper bitrates for the selected cameras (two or three and not be forced to cover the navigation window). In this case, no representation can be downloaded for the current video segment. To fairly compare with the other logics without this rebuffering event, we define the distortion to be 1 for this segment.

Finally, in order to observe the difference of the proposed optimal algorithm in the cases of the "Ad-hoc Set" and the "Optimized Set" at the server side, we define $\mathcal{L}3$ under the same storage constraint per video of 18Mbps. The camera view set is $\{1, 3, 5, 7, 10\}$ and the encoding bitrate set per view is $\{0.1, 0.3, 0.5, 2.7\}$(Mbps). The results of both cases for the "High-interactivity Scenario" are shown in Fig. 14. As expected, the case of the user having the "Optimized Set" achieves substantial improvements compared to the case having $\mathcal{L}3$. It is worth noting that, since the same bitrate set are allocated for all the views in the "Ad-hoc Set", there is a limitation for the maximum of overall bitrate that can be selected (e.g., only 13.5Mbps in $\mathcal{L}3$ when selecting 5 views at 2.7Mbps). Thus, as the bandwidth constraint is increased, the case having the "Optimized Set" obtains larger benefits. Overall, a better quality is achieved by our adaptation logic when the representation set is optimized at the server. This is because the right tradeoff between coding and synthesis artifacts is found at both the server and client sides in our designing architecture.

In summary, the above results under very different simulated settings have shown that the navigation distortion can be reduced for clients in interactive MV systems when the optimal representation set is selected following our joint optimization logic. This shows the importance of taking into account the content characteristics, bandwidth constraints, the users interactivity and representation sets available at the server when determining the content to be downloaded by the client.

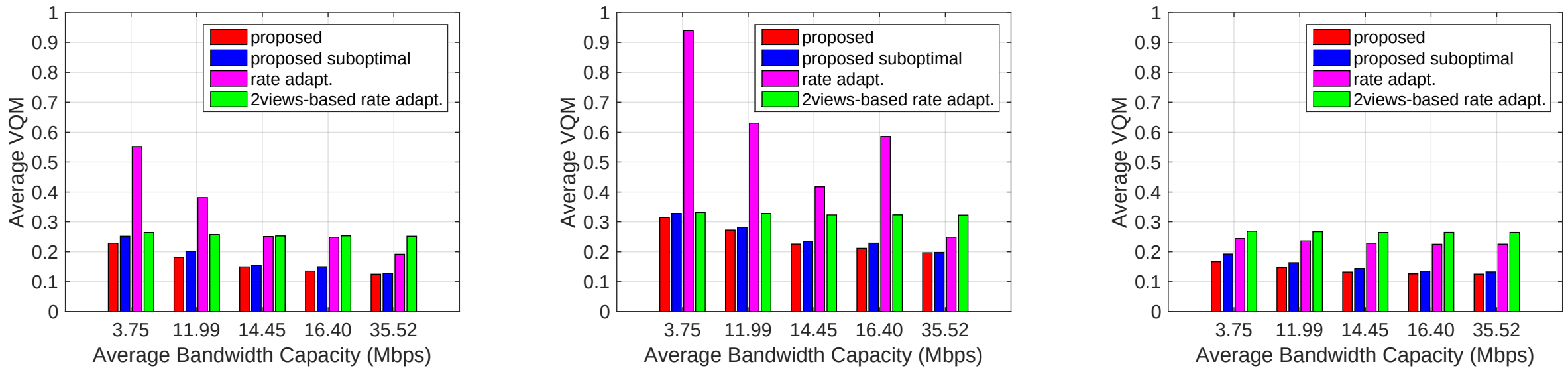

(a) *Dancer* (b) *Shark* (c) *Hall*

Fig. 13. Distortion comparison of different algorithms based on the "Optimized Set" for the "High-interactivity Scenario". (a) *Dancer* in the case of *uniform*, (b) *Shark* in the case of *non-uniform* with $p_n = 0.3$, and (c) *Hall* in the case of *non-uniform* with $p_n = 0.6$.

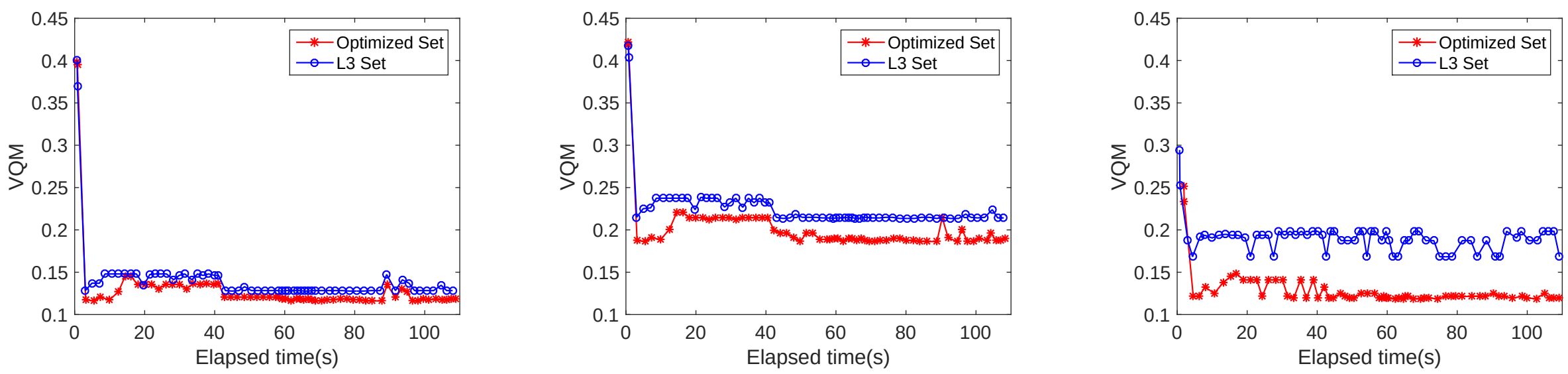

(a) *Dancer* (b) *Shark* (c) *Hall*

Fig. 14. Distortion comparison of proposed optimal algorithm using the "Optimized Set" or $\mathcal{L}3$ in the case of "High-interactivity Scenario" with the *non-uniform* navigation with $p_n = 0.6$ and a specific channel trace in Fig. 8.

## VII. Conclusion

This paper has studied a navigation-aware adaptation logic optimization problem for interactive free viewpoint video systems that is able to minimize both the navigation distortion and the view-switching delay. To the best of our knowledge, this is the first work to provide a formal optimization of HAS-client controller for adaptive MV streaming that jointly selects the best anchor views subsets and the corresponding encoding bitrates. Based on this optimization problem, we provide an optimal solution based on DP algorithm with polynomial time complexity, and an approximate algorithm with effective performance to further reduce computational complexity in practice. Our algorithm properly takes into account both video content characteristics and user interactivity level, and outperforms competitor algorithms under very diverse scenarios. We show that it is necessary to find the proper tradeoff between view quality and number of reference views in constrained-resource networks. Our future work will study the online adaptation strategy for HAS-client adaptive MV streaming, which learns online the temporal system evolution, such as dynamics of networks and navigation.

## Appendix A
## Quality comparison in terms of PSNR and SSIM

We provide here complementary results to Section VI.A. To further evaluate the performance, in Fig. 15, we offer the quality comparison in terms of PSNR and SSIM for the scenarios used in Fig. 4 (a). It can be observed that, similarly, both of the proposed optimal and suboptimal algorithms always outperform the baseline ones. Gains up to 3.98 dB and 0.12 in terms of PSNR and SSIM respectively are achieved by our optimal algorithm with respect to "view adaptation" algorithm for "Shark", and gains up to 1.8 dB and 0.01 are achieved with respect to "2views-based rate adaptation" for "Hall". Moreover, our suboptimal algorithm can obtain gains up to 3.37 dB and 0.11 for "Shark", and 1.78 dB and 0.01 for "Hall". It is shown again the necessary by jointly optimizing the set of camera views and their encoding bitrates for MV adaptive streaming.

## Appendix B
## Distortion comparison by considering quality variations

The complementary results taking account quality variations to Section VI.B are provided here. We evaluate the performance by $D_t(\mathcal{L}_d, w) + \beta(D_t(\mathcal{L}_d, w) - D_{t-1}(\mathcal{L}_d, w))$ with $\beta = 1$, and make a comparison over time in Fig. 16. We can notice that, for three video sequences, our optimal algorithm still outperforms all baselines and achieves the best navigation quality over time. In particular, for "Dancer", reduction up to 0.04, 0.07 and 0.07 in terms of average distortion over time are achieved with respect to "view adaptation", "rate adaptation" and "2views-based rate adaptation" logic, respectively. Similarly, the proposed suboptimal algorithm gets the gains up to 0.02, 0.05 and 0.05 for "Dancer". For "Hall", we recall the fact that joint coding is more efficient in the case of low bandwidth (e.g., from 20s to 40s).

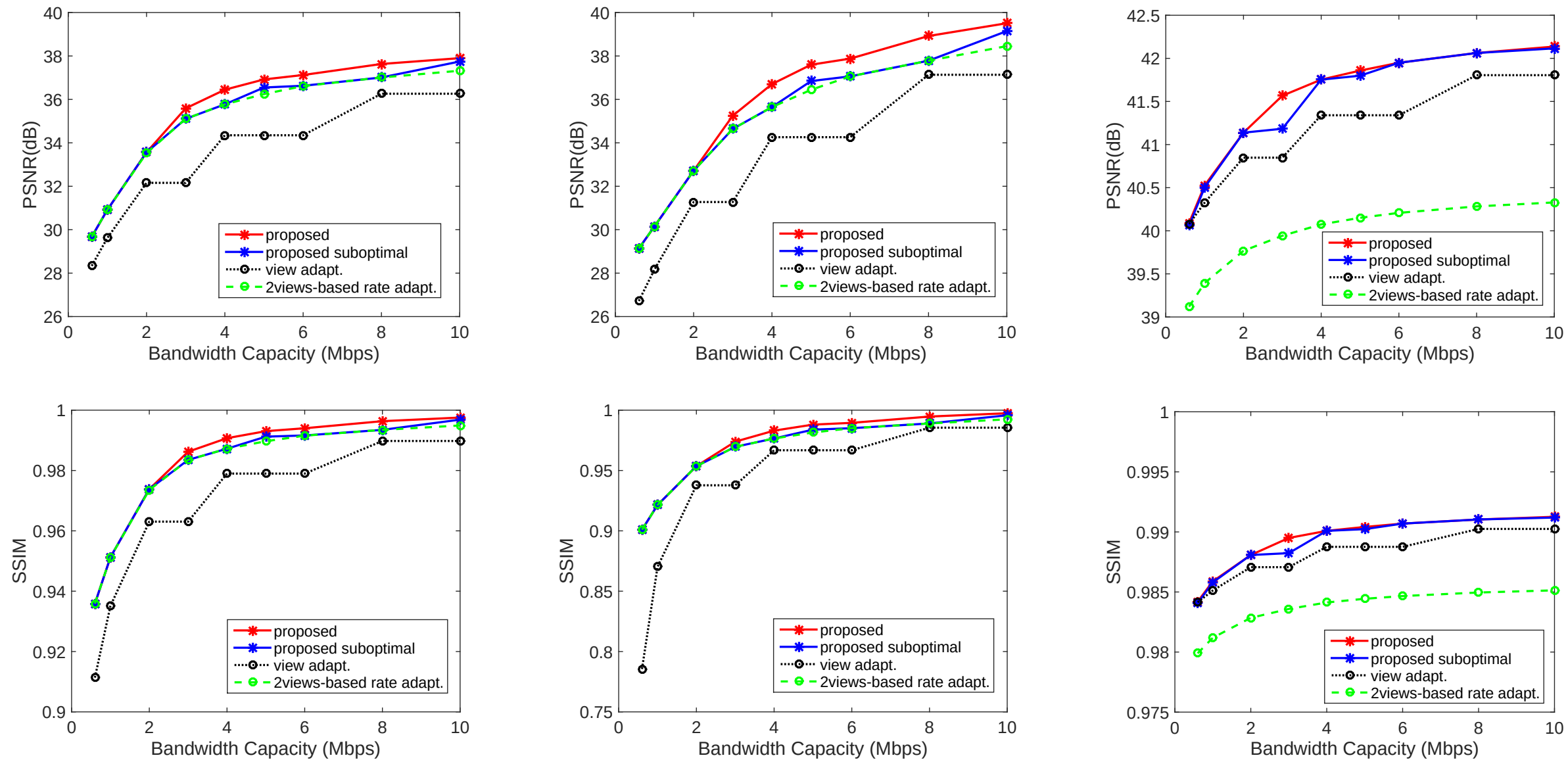

(a) *Dancer* (b) *Shark* (c) *Hall*

Fig. 15. Quality comparison in terms of PSNR (above) and SSIM (bottom) in the case of a client having navigation window $w = [5.5, 6.5]$ using $\mathcal{L}1$ for "Low-interactivity Scenario".

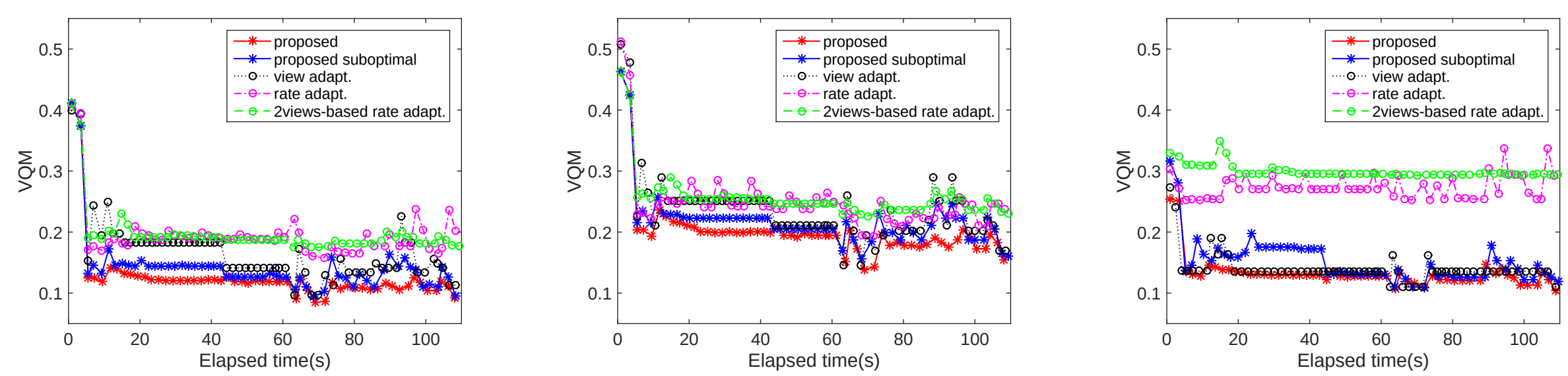

(a) *Dancer* (b) *Shark* (c) *Hall*

Fig. 16. Distortion comparison by considering the quality fluctuations over time in the case of *non-uniform* navigation model with $p_n = 0.6$ and a specific channel trace in Fig. 8 using $\mathcal{L}1$ for the "High-interactivity Scenario".

## Appendix C
## Comparison of Buffer Occupancy

We provide here complementary results to Section VI.B and look at the buffer occupancy at the client. To better understand the behavior of the realistic system, we depict the playout buffer sizes over time of all algorithms in the case of *non-uniform* navigation model with $p_n = 0.6$ in Fig. 17. We can note that our video player tries to maintain seamless playback over time without the occurrence of re-buffering events. This is because having the channel bandwidth as a constraint and an adaptation logic in realistic scenarios can effectively prevent the re-buffering in general, which is caused by downloading more than what the channel capacity can afford.

## Appendix D
## The Server-Side Optimized Set

We provide here more details about the optimized representation sets used in our experiments. We mostly follow the algorithm proposed in [22] to select the optimal representation sets for storage. In the case of "Low-interactivity Scenario", all users have a navigation window $w = [1.5, 9.5]$ for all video contents, while the experienced bandwidth per user follows a different channel trace. In particular, different channel traces represent different types of users, where each type of user has a bandwidth constraint averaged over one channel trace. The storage constraint of the server per video is fixed at 18Mbps. We finally define our representations with 15 types of users, namely, 3 types of videos $\times$ 5 possible bandwidth capacities (5 average bandwidth constraints, each is averaged over a channel trace). These 5 channel traces is the same ones in Fig. 11. In Fig. 18 (a), we show the optimal representations achieved by the joint optimization for three videos to be stored at the server. It can be noticed that, different videos have diverse

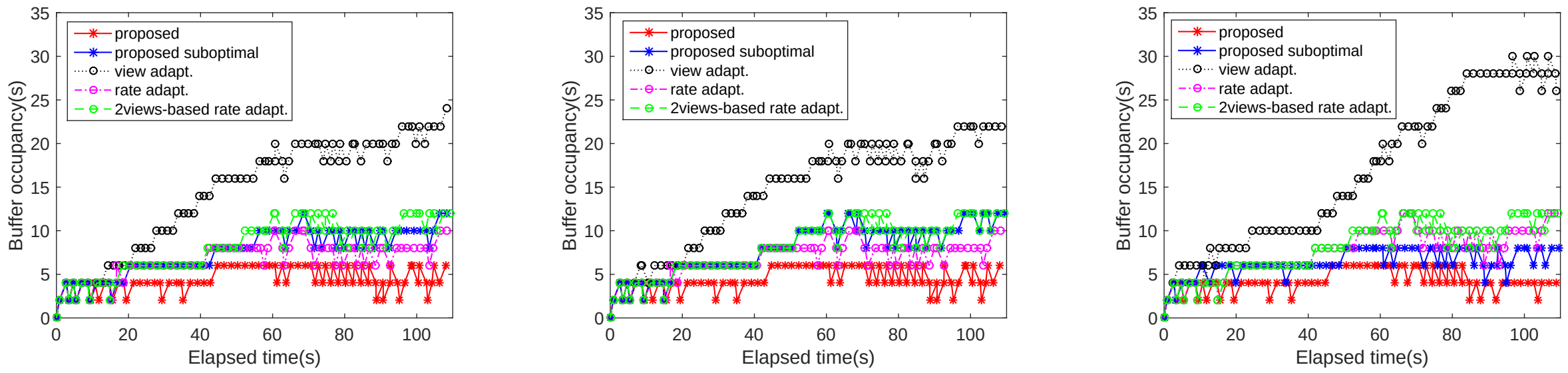

(a) *Dancer* (b) *Shark* (c) *Hall*

Fig. 17. Buffer comparison over time in the case of *non-uniform* navigation model with $p_n = 0.6$ and a specific channel trace in Fig. 8 using $\mathcal{L}1$ for the "High-interactivity Scenario".

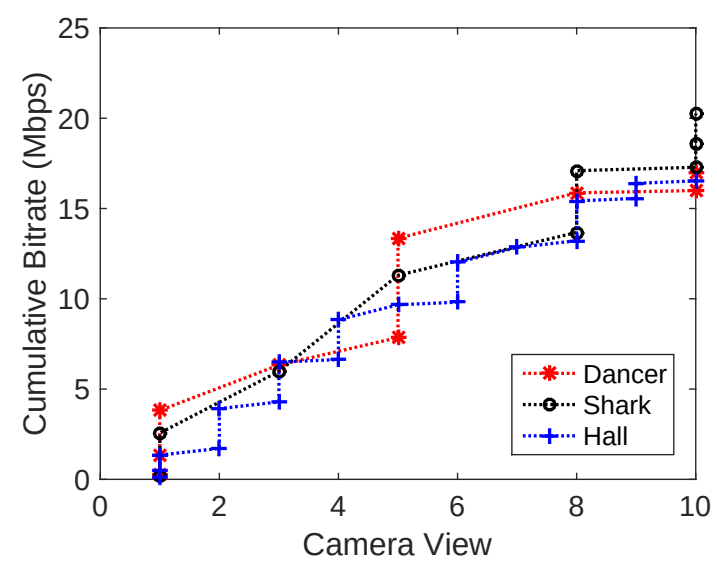

(a) "Low-interactivity Scenario"

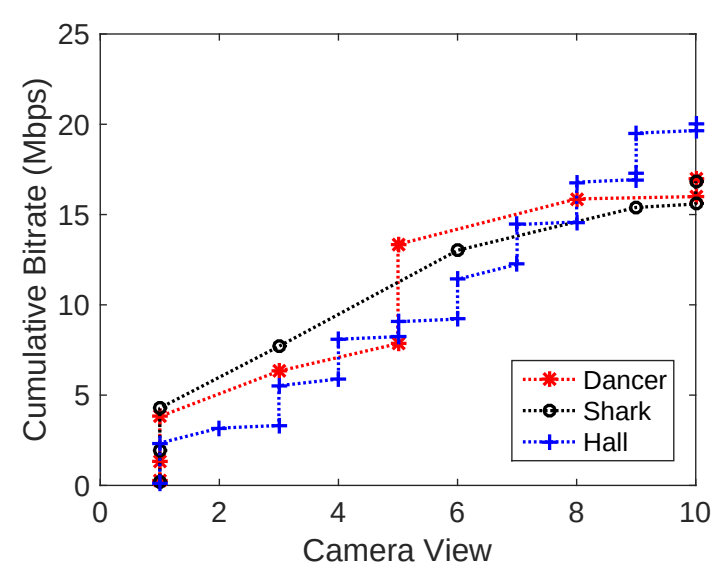

(b) "High-interactivity Scenario"

Fig. 18. The optimized representation sets at the server for different scenarios with the storage constraint per video of 18Mbps.

optimal sets, and different camera views are allocated unequal encoding bitrates as well.

In the case of "High-interactivity Scenario", we consider the probability of navigation window following a Normal random variable with variance $\sigma^2 = \{10, 150, 3000\}$ for respectively "*Dancer*", "*Shark*" and "*Hall*" as described in [22]. Similarly, 5 channel traces in Fig. 11 are experienced by different users, while 3 navigation windows are considered to change over time. The storage constraint of the server per video is set at 18Mbps as well. As a result, we define 45 types of users (3 types of videos $\times$ 3 possible navigation windows per video $\times$ 5 possible bandwidth capacities). The optimal representation set at the server achieved by the joint optimization for three sequences is illustrated in Fig. 18 (b).

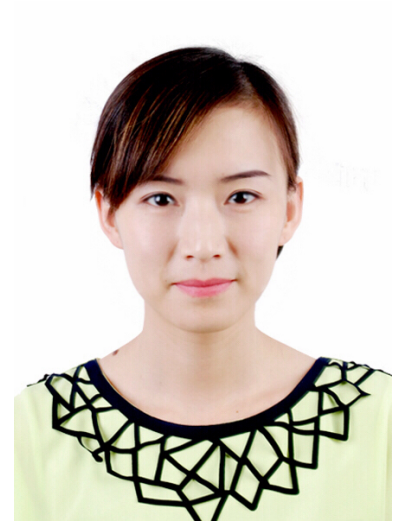

**Xue Zhang** is currently working toward the Ph.D. degree in Signal and Information Processing at Beijing Jiaotong University (BJTU), Beijing, China.

From October 2015 to September 2017, she was a Visiting Ph.D. student with the Signal Processing Laboratory (LTS4), Swiss Federal Institute of Technology (EPFL), Lausanne, Switzerland. From January 2018 to April 2018, she was a Visiting Ph.D. student with the National Institute of Informatics (NII), Tokyo, Japan. Her current research interests include image/video coding, 3D video processing, and interactive multimedia systems.

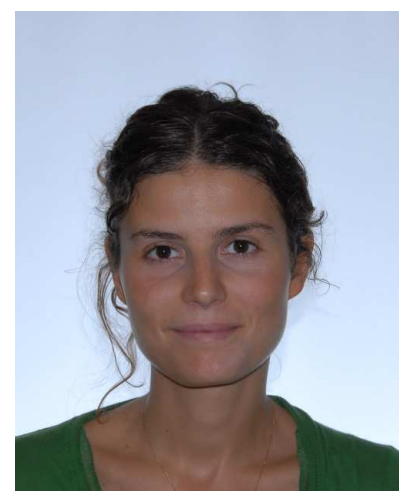

**Laura Toni** (S'06-M'09) received the M.S. and Ph.D. degrees, both in electrical engineering, from the University of Bologna, Italy, in 2005 and 2009, respectively. In 2007, she was a visiting scholar at the University of California at San Diego (UCSD), CA, and since 2009, she has been a frequent visitor to the UCSD, working on media coding and streaming technologies.

Between 2009 and 2011, she worked at the Tele-Robotics and Application (TERA) department at the Italian Institute of Technology (IIT), investigating wireless sensor networks for robotics applications. In 2012, she was a Post-doctoral fellow at the UCSD and between 2013 and 2016 she was a Post-doctoral fellow in the Signal Processing Laboratory (LTS4) at the Swiss Federal Institute of Technology (EPFL), Switzerland. Since July 2016, she has been appointed as Lecturer in the Electronic and Electrical Engineering Department of University College London (UCL), UK.

Her research mainly involves interactive multimedia systems, decision-making strategies under uncertainty, large-scale signal processing and communications. She received the UCL Future Leadership Award in 2016, the ACM best 10% paper award in 2013, and the IEEE/IFIP best paper award in 2012.

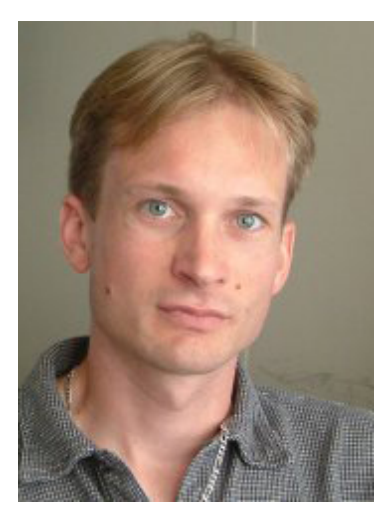

**Pascal Frossard** (S'96-M'01-SM'04-F'18) received the M.S. and Ph.D. degrees, both in electrical engineering, from the Swiss Federal Institute of Technology (EPFL), Lausanne, Switzerland, in 1997 and 2000, respectively. Between 2001 and 2003, he was a member of the research staff at the IBM T. J. Watson Research Center, Yorktown Heights, NY, where he worked on media coding and streaming technologies. Since 2003, he has been a faculty at EPFL, where he heads the Signal Processing Laboratory (LTS4). His research interests include graph signal processing, image representation and coding, visual information analysis, and distributed signal processing and communications.

Dr. Frossard has been the General Chair of IEEE ICME 2002 and Packet Video 2007. He has been the Technical Program Chair of IEEE ICIP 2014 and EUSIPCO 2008, and a member of the organizing or technical program committees of numerous conferences. He has been a senior Area Editor of the IEEE TRANSACTIONS ON SIGNAL PROCESSING (2015-), an Associate Editor of the IEEE TRANSACTIONS ON BIG DATA (2015-), IEEE TRANSACTIONS ON IMAGE PROCESSING (2010-2013), the IEEE TRANSACTIONS ON MULTIMEDIA (2004-2012), and the IEEE TRANSACTIONS ON CIRCUITS AND SYSTEMS FOR VIDEO TECHNOLOGY (2006-2011). He is an elected member of the IEEE Multimedia Signal Processing Technical Committee (2004-2007, 2016-), the IEEE Visual Signal Processing and Communications Technical Committee (2006-) and the IEEE Multimedia Systems and Applications Technical Committee (2005-). He has served as Chair of the IEEE Image, Video and Multidimensional Signal Processing Technical Committee (2014-2015), and Steering Committee Chair (2012-2014) and Vice-Chair (2004-2006) of the IEEE Multimedia Communications Technical Committee. He received the Swiss NSF Professorship Award in 2003, the IBM Faculty Award in 2005, the IBM Exploratory Stream Analytics Innovation Award in 2008, the Google Faculty Award 2017, the IEEE Transactions on Multimedia Best Paper Award in 2011, and the IEEE Signal Processing Magazine Best Paper Award 2016. He is a Fellow of the IEEE.

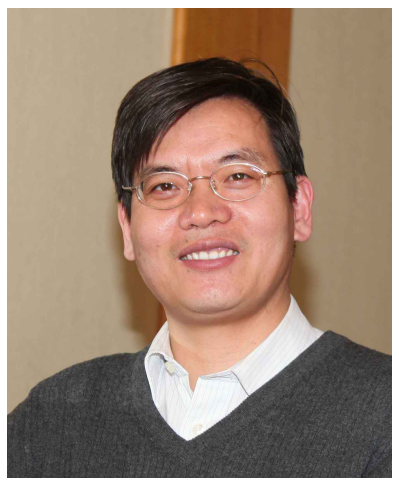

**Yao Zhao** (M'06-SM'12) received the B.S. degree from Fuzhou University, Fuzhou, China, in 1989, and the M.E. degree from Southeast University, Nanjing, China, in 1992, both from the Radio Engineering Department, and the Ph.D. degree from the Institute of Information Science, Beijing Jiaotong University (BJTU), Beijing, China, in 1996.

He became an Associate Professor at BJTU in 1998 and became a Professor in 2001. From 2001 to 2002, he was a Senior Research Fellow with the Information and Communication Theory Group, Faculty of Information Technology and Systems, Delft University of Technology, Delft, The Netherlands. In October 2015, he visited the Swiss Federal Institute of Technology, Lausanne, Switzerland. From December 2017 to March 2018, he visited the University of Southern California. He is currently the Director of the Institute of Information Science, BJTU. His current research interests include image/video coding, digital watermarking and forensics, and video analysis and understanding.

Dr. Zhao serves on the Editorial Boards of several international journals, including as an Associate Editor of the IEEE TRANSACTIONS ON CYBERNETICS, an Associate Editor of the IEEE SIGNAL PROCESSING LETTERS, and an Area Editor of *Signal Processing: Image Communication*. He was named a Distinguished Young Scholar by the National Science Foundation of China in 2010, and was elected as a Chang Jiang Scholar of Ministry of Education of China in 2013. He is a Fellow of the IET.

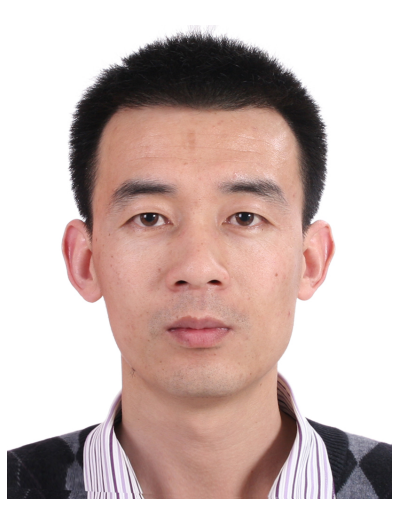

**Chunyu Lin** was born in Liaoning Province, China. He received the Doctor degree from Beijing Jiaotong University, Beijing, China, in 2011.

From 2009 to 2010, he was a Visiting Researcher at the ICT Group, Delft University of Technology, Delft, The Netherlands. From 2011 to 2012, he was a Post-Doctoral Researcher with the Multimedia Laboratory, Gent University, Gent, Belgium. He is currently an associate professor in Beijing Jiaotong University. His research interests include image/video compression and robust transmission, 2D-to-3D conversion, 3-D video coding and virtual reality video processing.